\documentclass[twocolumn]{aastex63}

\usepackage{amsmath}
\usepackage{lineno}
\usepackage{CJK}
\usepackage{amsbsy}
\usepackage{multirow}
\usepackage{bm}
\usepackage{ulem}
\usepackage{underscore}
\received{XXX}
\revised{YYY}
\accepted{ZZZ}
\published{HHH}
\submitjournal{ApJ}

\shorttitle{NIHAO-RiNG}
\shortauthors{H.Z. Chen et al.}
\begin{document}
\begin{CJK*}{UTF8}{gbsn}

\title{NIHAO-RiNG: A Comparison of Simulated Disc Galaxies from GASOLINE and GIZMO}

\correspondingauthor{Hou-Zun Chen}
%\email{houzunchen@pmo.ac.cn}
\email{chenhz_zju@zju.edu.cn}
\correspondingauthor{Xi Kang}
\email{kangxi@zju.edu.cn}

\author[0000-0001-8426-9493]{Hou-Zun Chen (陈厚尊)}
\affiliation{Institute for Astronomy, the School of Physics, Zhejiang University, Hangzhou 310027, China}
\affiliation{Purple Mountain Observatory, 10 Yuan Hua Road, Nanjing 210034, China}
\affiliation{School of Astronomy and Space Sciences, University of Science and Technology of China, Hefei 230026, China}

\author[0000-0002-5458-4254]{Xi Kang (康熙)}
\affiliation{Institute for Astronomy, the School of Physics, Zhejiang University, Hangzhou 310027, China}
\affiliation{Purple Mountain Observatory, 10 Yuan Hua Road, Nanjing 210034, China}
\affiliation{Center for Cosmology and Computational Astrophysics, Zhejiang University, Hangzhou 310027, China}

\author[0000-0002-8171-6507]{Andrea V. Macci\`o}
\affiliation{New York University Abu Dhabi, PO Box 129188, Saadiyat Island, Abu Dhabi, United Arab Emirates}
\affiliation{Center for Astrophysics and Space Science (CASS), New York University Abu Dhabi}
\affiliation{Max-Planck-Institut f\"ur Astronomie, K\"onigstuhl 17, 69117 Heidelberg, Germany}
\author[0000-0003-2027-399X]{Tobias Buck}
\affiliation{Universit\"at Heidelberg, Interdisziplin\"ares Zentrum f\"ur Wissenschaftliches Rechnen, Im Neuenheimer Feld 205, 69120 Heidelberg, Germany}
\affiliation{Universit\"at Heidelberg, Zentrum f\"ur Astronomie, Institut f\"ur Theoretische Astrophysik, Albert-Ueberle-Straße 2, 69120 Heidelberg, Germany}

\author[0000-0001-8531-9536]{Renyue Cen}
\affiliation{Center for Cosmology and Computational Astrophysics, Zhejiang University, Hangzhou 310027, China}
\affiliation{Institute for Astronomy, the School of Physics, Zhejiang University, Hangzhou 310027, China}

\begin{abstract}

We utilize the public \texttt{GIZMO} code to re-simulate 12 galaxies selected from the NIHAO simulation suite which were run with the \texttt{GASOLINE} code, then compare their properties. We find that while both codes with the same initial conditions and large-scale environments can successfully produce similar galactic discs in Milky Way-mass systems, yet significant differences are still seen in many aspects, particularly the circum-galactic medium (CGM) environment they reside in. Specifically, the thermal feedback recipe used in \texttt{GASOLINE} results in ubiquitous long-term large-scale outflows, primarily driven by high-density hot inter-stellar medium (ISM) from the galaxy center, preventing the inter-galactic medium (IGM) from falling efficiently. Recycled gas and inflows in CGM appear at $10^{4\sim5}\rm K$, playing a crucial role in the formation of cold discs in the NIHAO simulations. In contrast, disc galaxies simulated by \texttt{GIZMO} do not exhibit prominent outflows at low redshifts, but instead display quasi-virialized hot gas halos that arise from the interaction between inflows and feedback-driven outflows. Therefore, the origins of mass and angular momentum of the cold disc in the two simulations are quite different, even though the final morphology of corresponding galaxies are both disky. The differences in the distribution of CGM gas are mainly due to different feedback models implemented in the two codes, future observations of CGM could provide valuable insight into the physics governing baryon cycle in galaxies.

\end{abstract}


\section{Introduction} \label{sec:1}

Ab initio simulation of disc dominated galaxies has turned out to be a difficult task since the 1970s when simulators began to tackle the formation of Milky Way (MW) like systems by applying relevant physical processes to applicable cosmological initial conditions \citep{1977ApJ...215..483B, 1977MNRAS.179..541R}. These physical processes include pure gravitational interactions (i.e., formation of dark matter halos and their growth and mergers) and hydro/thermal-dynamics processes, such as gas heating, cooling, dissipation, star formation and feedback etc. It has been realized that the conservation of angular momentum of cooling gas within the dark matter halos is key to form disc galaxy with a flat rotation curve as observed \citep{1998MNRAS.295..319M}.  However, the over-cooling problem, i.e. gas at high redshifts will fall into high dense regions and produce too many stars, making the formation of disc dominated galaxies difficult in simulations \citep{1995MNRAS.275...56N, 1997ApJ...478...13N}. It indicates that some feedback processes from young stellar populations are essential since they could eject large amounts of gas and thus prevent the over-efficient star formations \citep{2013MNRAS.428..129S,2013ApJ...770...25A}.

In the past decades, various sub-grid feedback processes have been proposed in hydro-dynamical simulations to describe the multi-phase inter-stellar medium (ISM) and circum-galactic medium (CGM). Models experimented in hydro-simulations can be broadly divided into several classes, e.g.， delayed cooling model \citep{2006MNRAS.373.1074S}, stochastic thermal feedback \citep{2012MNRAS.426..140D}, two-phase approach \citep{2006MNRAS.371.1125S}, wind feedback \citep{2003MNRAS.339..289S} etc. With strong feedback from stellar evolution,  disc galaxies with more realistic properties could be reproduced by different simulations \citep[for a review see][]{2017ARA&A..55...59N}, but the detailed results of simulated disc galaxies are usually sensitive to sub-grid physics in simulations, and it severely limits the predictive power of simulation. Therefore, it is important to compare how the simulation methods and implemented physics affect the outcomes of simulations, as have done by projects such as Aquila \citep{2012MNRAS.423.1726S} and  Agora \citep{2014ApJS..210...14K} to compare the results from various cosmological simulations. \citet{2023ApJ...950..132H} also presented a suite of high-resolution simulations of an isolated dwarf galaxy to compare the performance of different simulation code GIZMO \citep{2015MNRAS.450...53H}, GADGET \citep{2005MNRAS.364.1105S}, AREPO \citep{2016MNRAS.455.1134P} and RAMSES \citep{2002A&A...385..337T}.

More recently, a supernova feedback scheme with both energy and momentum deposition that ensures to correctly capture the terminal momentum, regardless of the numerical simulation resolution, have been implemented in cosmological simulations \citep{2014ApJ...788..121K}. In the limit of good resolutions where the cooling radius is resolved, the scheme correctly follow the evolution of Sedov-Taylor blastwaves with cooling from the earliest phase resolved, as nature does. In the limit of poor resolutions where the cooling radius is not adequately resolved, the scheme still yields the correct momentum and thermal energy at the resolution scale at the corresponding evolutionary phase, which may be the late Sedov-Taylor or the momentum conserving phase. It is demonstrated in \cite{2015MNRAS.451.2900K} that different implementations of supernova feedback can result in very different outcomes, such as the stellar mass-metallicity relation.

In the last years, hydro-dynamical simulations, such as NIHAO \citep{2015MNRAS.454...83W}, FIRE \citep{2018MNRAS.480..800H}, EAGLE \citep{2015MNRAS.446..521S} and IllustrisTNG \citep{2018MNRAS.473.4077P} have achieved great success in reproducing various properties of local galaxies, such as stellar mass, disc size, and metal distribution. However, it is found that their predictions regarding the CGM exhibit significant diversity \citep[for more details see][]{2021MNRAS.504.5131L, 2022MNRAS.514.3113K, 2022ApJ...936L..15C}. As galaxy properties depend strongly on its formation history, even for one single cosmological simulation, the properties of CGM can vary on a galaxy-to-galaxy basis. Therefore, before one can draw conclusions on how galaxy population differs between different cosmological simulations, it is important to check how different simulation code behave for one single galaxy with a specific assembly history. This could be done by simulating the same galaxy (the same initial conditions) with different codes.

In this study, we aim to investigate the differences in predictions and the impact of different treatments of sub-grid physics on galaxy properties. To achieve this, we use the public \texttt{GIZMO} code to re-simulate a few galaxies whose initial conditions are from the NIHAO sample. We refer to this comparison as RiNG project (short for \textbf{R}e-s\textbf{i}mulated \textbf{N}IHAO \textbf{G}alaxy), which involves utilizing different codes, \texttt{GASOLINE} and \texttt{GIZMO} in this work, simulating galaxies with identical initial conditions. Through this comparison, we examine the properties of CGM, cold disc and feedback gas (see the definition in Section \ref{sec:2.3}), and gain insights into how different sub-grid physics treatments implemented in the two codes impact those properties. 

This paper is organized as follows: we introduce the basic parameters and numerical methods in Section \ref{sec:2}. In Section \ref{sec:3} we show the simulated galaxy properties such as morphology, phase diagram, profiles, baryon fraction, feedback gas, star formation histories etc. Conclusions and discussions of this work will be given in Section \ref{sec:4}.

\section{Methods and Simulations} \label{sec:2}

\begin{table*}[htbp]
\caption{Basic properties of the 12 simulated galaxies in this work. Columns from left to right: system name and box size in $\mathrm{Mpc}/h$, particle mass of dark matter and baryon, force softenings of dark matter and baryon, system virial mass derived by \texttt{GASOLINE} and \texttt{GIZMO} at $z=0$, gas mass derived by \texttt{GASOLINE} and \texttt{GIZMO} at $z=0$, whether the galaxy is disc dominated at $z=0$ (see the mock optical images in face-on and edge-on view in Fig. \ref{fig:mocks}). Numbers in brackets denote the spin parameter \citep{1969ApJ...155..393P} of the halo in which the galaxy is embedded. Note that the virial mass derived from two runs is usually different from the halo mass derived from the dark matter only run.}
\label{tab:tab1}
\centering
\begin{tabular}{c|c|c|c|c|c|c|c|c}
\hline\hline
 \multirow{2}*{system (box)} & \multirow{2}*{$m_{\mathrm{dm}}/m_{\mathrm{b}}$ ($10^5 \mathrm{M_{\odot}}$)} & \multirow{2}*{$\epsilon_{\mathrm{dm}}/\epsilon_{\mathrm{b}}\left(\mathrm{pc}\right)$} & \multicolumn{2}{c|}{virial mass ($\mathrm{M_{\odot}}$)} & \multicolumn{2}{c|}{gas mass ($\mathrm{M_{\odot}}$)} & \multicolumn{2}{c}{disc or not ($\lambda$)}\\ \cline{4-9}
 & & & \texttt{GASOLINE} & \texttt{GIZMO} & \texttt{GASOLINE} & \texttt{GIZMO} & \texttt{GASOLINE} & \texttt{GIZMO} \\ \hline
g3.54e09 (15) & 0.0339 / 0.00618 & 38.6 / 16.6 & $3.95\times10^9$ & $3.97\times10^9$ & $2.65\times10^7$ & $8.07\times10^6$ & $\times\left(0.019\right)$ & $\times\left(0.019\right)$\\
g2.94e10 (20) & 0.190 / 0.0347 & 66.7 / 26.7 & $3.22\times10^{10}$ & $3.37\times10^{10}$ & $5.72\times10^8$ & $4.15\times10^8$ & $\times\left(0.051\right)$ & $\times\left(0.055\right)$\\
g7.12e10 (60) & 0.643 / 0.117 & 267 / 113 & $2.88\times10^{11}$ & $3.73\times10^{11}$ & $5.37\times10^9$ & $1.09\times10^{10}$ & $\times\left(0.054\right)$ & $\surd\left(0.037\right)$ \\
g8.89e10 (60) & 2.169 / 0.396 & 267 / 113 & $9.22\times10^{10}$ & $1.03\times10^{11}$ & $5.60\times10^9$ & $5.67\times10^9$ & $\surd\left(0.064\right)$ & $\surd\left(0.066\right)$ \\
g1.52e11 (60) & 2.169 / 0.396 & 267 / 113 & $1.57\times10^{11}$ & $1.66\times10^{11}$ & $1.38\times10^{10}$ & $1.11\times10^{10}$ & $\surd\left(0.081\right)$ & $\surd\left(0.081\right)$ \\
g5.36e11 (60) & 2.169 / 0.396 & 267 / 113 & $6.89\times10^{11}$ & $6.82\times10^{11}$ & $4.96\times10^{10}$ & $5.18\times10^{10}$ & $\times\left(0.099\right)$ & $\surd\left(0.061\right)$ \\
g6.96e11 (60) & 1.523 / 0.278 & 267 / 113 & $6.76\times10^{11}$ & $8.20\times10^{11}$ & $4.79\times10^{10}$ & $7.50\times10^{10}$ & $\surd\left(0.060\right)$ & $\surd\left(0.037\right)$ \\
g7.08e11 (60) & 1.110 / 0.203 & 267 / 113 & $5.50\times10^{11}$ & $6.23\times10^{11}$ & $3.74\times10^{10}$ & $4.58\times10^{10}$ & $\surd\left(0.079\right)$ & $\surd\left(0.077\right)$ \\
g7.55e11 (60) & 1.523 / 0.278 & 267 / 113 & $8.54\times10^{11}$ & $9.57\times10^{11}$ & $6.80\times10^{10}$ & $8.83\times10^{10}$ & $\surd\left(0.053\right)$ & $\surd\left(0.047\right)$ \\
g8.26e11 (60) & 2.169 / 0.396 & 267 / 113 & $9.10\times10^{11}$ & $1.01\times10^{12}$ & $6.09\times10^{10}$ & $8.00\times10^{10}$ & $\surd\left(0.078\right)$ & $\surd\left(0.075\right)$ \\
g1.12e12 (60) & 1.523 / 0.278 & 200 / 66.7 & $1.28\times10^{12}$ &  $1.41\times10^{12}$ & $7.93\times10^{10}$ & $1.26\times10^{11}$ & $\times\left(0.024\right)$ & $\times\left(0.020\right)$\\
g2.79e12 (60) & 5.141 / 0.938 & 300 / 100 & $3.13\times10^{12}$ & $3.93\times10^{12}$ & $1.85\times10^{11}$ & $2.97\times10^{11}$ & $\surd\left(0.047\right)$ & $\surd\left(0.049\right)$\\ \hline\hline
\end{tabular}
\end{table*}

\subsection{NIHAO Recipe and Sample Selection}\label{sec:2.0}

The NIHAO project \cite[Numerical Investigation of a Hundred Astronomical Object from][which will be denoted as NIHAO-I hereafter]{2015MNRAS.454...83W} and its high-resolution version, the NIHAO-UHD simulations \citep{2020MNRAS.491.3461B}, were designed to create a large sample of cosmological zoom-in simulations over a range of masses from dwarf $(M_{\rm halo}\sim5\times10^9\mathrm{M}_{\odot})$ to MW-mass $(M_{\rm halo}\sim2\times10^{12}\mathrm{M}_{\odot})$ with an unbiased sampling of the mass accretion histories. NIHAO suite (for convenience, we will use NIHAO suite as the collective term for both NIHAO-I and NIHAO-UHD) were performed using the \texttt{GASOLINE} code with modern Smoothed Particle Hydrodynamics (SPH) algorithm \citep{2004NewA....9..137W}. The fiducial runs adopt $T<1.5\times10^4\,\mathrm{K}$ and $n_{\mathrm{th}}>10.3\,\mathrm{cm}^{-3}$ as the temperature and density thresholds for star formation. The stars feed thermal energy back through blast-wave supernova feedback and ionizing feedback from massive stars prior to their explosion, i.e., ``early stellar feedback" \citep{2013MNRAS.428..129S}. The fiducial runs also set pre-supernova (pre-SN) feedback to $\epsilon_{\mathrm{ESF}}=13$ per cent of the total stellar flux, and a $30\mathrm{Myr}$ delayed cooling time for gas particles inside the blast region \citep{2015MNRAS.454...83W}. NIHAO-I and NIHAO-UHD simulations have established good agreement with observational results for the MW-mass galaxies and local disc galaxies from the SPARC sample \citep{2016AJ....152..157L,2016MNRAS.463L..69M,2018MNRAS.473.4392S,2020MNRAS.491.3461B}. 

In this work, we select a total of 12 galaxies, with half of them are MW-mass samples from the NIHAO-UHD, namely g6.96e11, g7.08e11, g7.55e11, g8.26e11, g1.12e12 and g2.79e12 (the system name denotes the halo mass of the corresponding dark matter only run), the other 6 galaxies, namely g3.54e09, g2.94e10, g7.12e10, g8.89e10, g1.52e11, g5.36e11, are low-mass systems from NIHAO-I. All these galaxies have the same the order of baryonic resolution ($m_{\rm b}\sim10^4\,\mathrm{M}_{\odot}$, see Table \ref{tab:tab1} for detail). We then re-simulate the 12 galaxies (the results referred to RiNG hereafter) using the public \texttt{GIZMO} code \citep{2015MNRAS.450...53H}, which is a massively-parallel, multi-method code with various physics modules. \texttt{GIZMO} is modified from \texttt{P-GADGET} \citep{2005MNRAS.364.1105S}, and is fully compatible with Gadget-type initial conditions or snapshots. In Table \ref{tab:tab1} we list the details of all galaxy sample.

The RiNG and NIHAO suite adopt the same cosmological parameters from the \cite{2014A&A...571A..16P}, namely: $\Omega_m = 0.3175$, $\Omega_\Lambda = 0.6825$, $\Omega_b = 0.049$, $H_0 = 67.1\,\mathrm{km\,s}^{-1}\mathrm{Mpc}^{-1}$ and $\sigma_8 = 0.8344$. As both simulations share the same initial conditions created with a modified version of the \texttt{GRAFIC2} package \cite[see][]{2015MNRAS.454...83W}, it is mainly the numerical method and baryonic physics, such as hydrodynamic solver, gas cooling and stellar feedback, that account for the different properties between corresponding samples. In the following section, we present the main difference of implementations in the two codes.

\subsection{Hydrodynamics and Cooling} \label{sec:2.1}

The hydrodynamic solver implemented in \texttt{GIZMO} is called the meshless finite-volume method (MFM), which is also the default hydrodynamic solver of the FIRE simulations \citep{2018MNRAS.480..800H}. MFM solves hydrodynamics using a mesh-free Lagrangian (fixed-mass) finite-volume Godunov method designed to capture advantages of both grid-based and particle-based methods \citep{2011MNRAS.414..129G}. More information on this solver can be found in \cite{2015MNRAS.450...53H}.

Gas cooling and heating processes we used here are coupled with an alternative non-equilibrium chemical (ion+atomic+molecular) network, namely the CHIMES module, which is mainly developed by \cite{2014MNRAS.440.3349R, 2014MNRAS.442.2780R}. This module solves a large molecular and ion network such as photo-ionization, photo-dissociation, cosmic ray ionization, dust grain physics, metal line cooling, $\mathrm{H}_2$ rovibrational cooling, molecule cooling, photo-heating, photoelectric heating etc. So one can trace predictive chemistry for species in dense ISM gas in much greater detail, but with much additional computational expense. As the focus of this work is not on the metallicity issues, to speed up the simulations, we only enable the network part that involves primordial elements. For the metal part, we simply trace 9 elements separately, i.e. C, N, O, Ne, Mg, Si, S, Ca, Fe \citep{2014MNRAS.445..581H}.
%, which relevant ionization states are tabulated from \texttt{CLOUDY} simulations \cite[from][]{2009ApJ...703.1416F}.

\subsection{Star Formation and Feedback} \label{sec:2.2}

\texttt{GIZMO} turns cold gas into stars when it is locally self-gravitating, self-shielding, Jeans unstable, and above certain density threshold ($n_{\mathrm{th}}>100\,\mathrm{cm}^{-3}$). Once a star particle is formed, it will inherit the metallicity and mass from the progenitor gas particle. The SN rate is inferred separately from type Ia and type II by assuming different event rates \cite[see APPENDIX A of FIRE-2 paper, i.e.][]{2018MNRAS.480..800H}. The smallest time step in our zoom-in simulation ($\Delta t$) is usually short enough ($10^{3\sim4}\,\rm yr$) to determine whether a single SN event happens in a star particle, which means that single SN explosion event can be treated explicitly. Here we employ the mechanical feedback model \citep{2014MNRAS.445..581H, 2018MNRAS.477.1578H} to tackle single SN event. For completeness we just give a glimpse of this feedback model here: for every single star particle, an event probability $p$ is determined for $\Delta t$, if a SN event happens, the appropriate ejecta mass, metallicity, energy, and momentum are deposited directly into the surrounding gas around the star particle. The algorithm for deposition is constructed to ensure machine-accurate conservation of mass, metallicity, energy, and momentum, while also ensuring that the ejecta are isotropic in the rest frame of star particle \citep{2018MNRAS.477.1578H}.

\subsection{Gas Components and Post-processing} \label{sec:2.3}

For simple consideration, the multi-phase gas in our sample is classified into several components by the distance to galaxy center ($r$) and temperature ($T$), i.e. ISM ($r<0.15R_{\mathrm{vir}}$), CGM ($0.15R_{\mathrm{vir}}<r<R_{\mathrm{vir}}$), intergalactic medium (IGM, $r>R_{\mathrm{vir}}$), as well as cold ($T<1.5\times10^4\,\mathrm K$), warm ($1.5\times10^4\,\mathrm K<T<10^5\mathrm K$), hot ($T>10^5\mathrm K$) and their combinations. Here $R_{\rm vir}$ denotes the virial radius identified by the \texttt{AHF} (see below). For each snapshot we also define feedback gas as that which initially resides in the cold ISM but is then moved to hot/warm ISM (thermal feedback dominated) or CGM (kinetic feedback dominated) at the next snapshot. Unless otherwise specified, the feedback gas particle mentioned in following sections is identified by this criterion.

In our work, main halos and substructures at each snapshot are identified by the hybrid halo finder \texttt{Amiga Halo Finder} \citep[\texttt{AHF}, ][]{2004MNRAS.351..399G, 2009ApJS..182..608K}, which is also used by the NIHAO suite. The virial mass of a halo is defined as the mass of all particles within a sphere containing $200$ times the cosmic critical matter density ($M_{\rm vir}$). The radius of this sphere is defined accordingly as the virial radius ($R_{\rm vir}$). We also made use of the \texttt{PYNBODY} \citep{2013ascl.soft05002P} package to filter and trace particles across snapshots.

\section{Results} \label{sec:3}

\begin{figure*}[htp]
\plotone{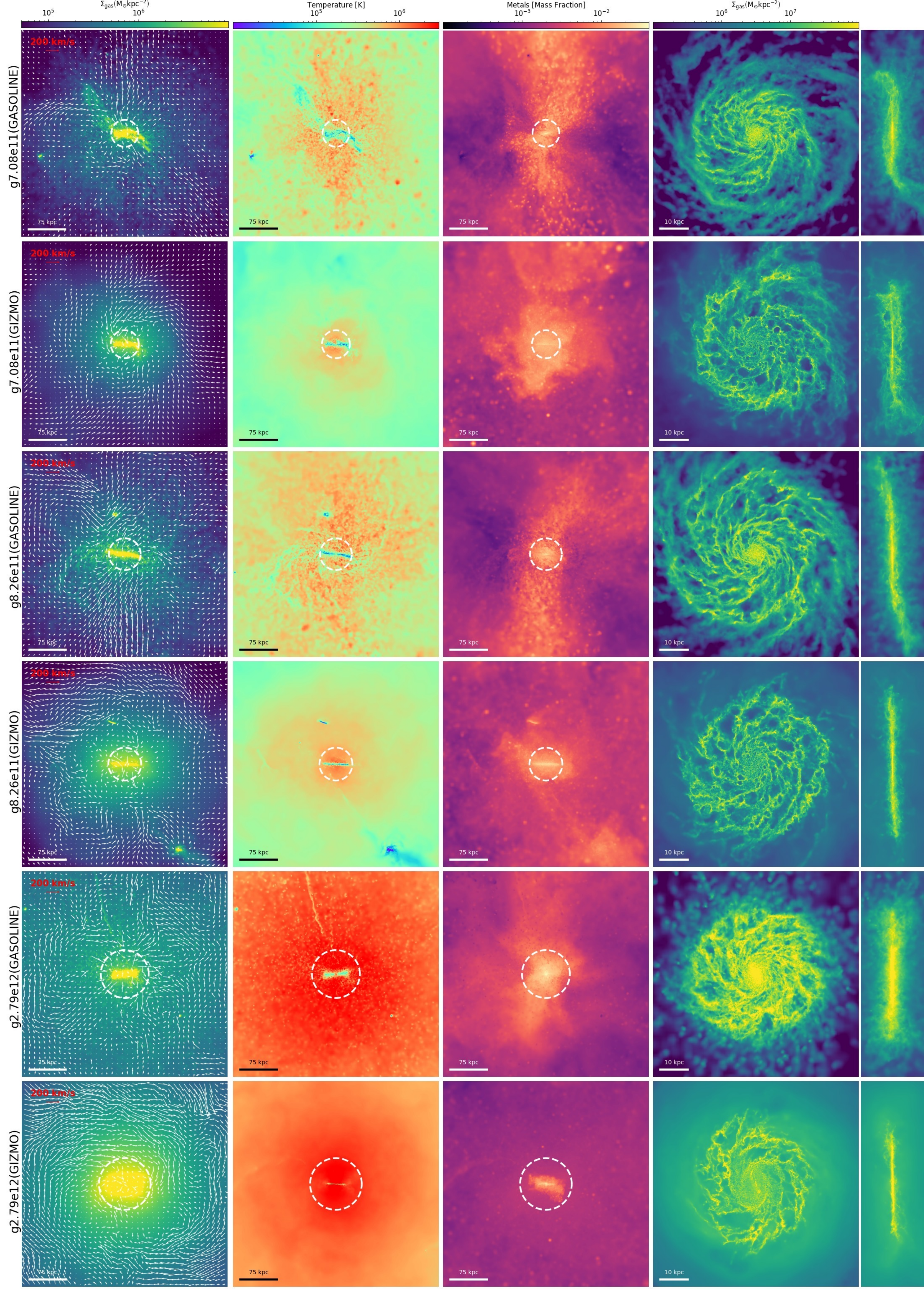}
\caption{Images of three disc galaxies (g7.08e11, g8.26e11, g2.79e12) simulated by \texttt{GASOLINE} and \texttt{GIZMO} at $z=0$. The dashed circle indicates the range of 15\% virial radius. System names and codes are labelled on the left. Columns from left to right: 2D density map of gas in edge-on view with pixelized velocity vectors overlaid on top, 2D temperature map averaged by gas density in edge-on view, 2D metallicity map averaged by gas density in edge-on view, 2D zoom-in density map of gas disc in face-on and edge-on view.}
\label{fig:fig1}
\end{figure*}

In this section we present our main findings. It is important to note that even for every single galaxy, the two codes produce different results to some extent. \textbf{In this paper we focus on MW-mass galaxies and present the main conclusions that are broadly applicable to the MW-mass disc galaxies studied here, unless otherwise specified in the text}. To begin, we show the comparison of three galaxies in Fig. \ref{fig:fig1}, highlighting the gas properties of the three disc galaxies in ascending order of virial mass: g7.08e11, g8.26e11, and g2.79e12. It is worth noting that the spin parameters of their host halos are consistent in both simulations (see the last two columns in Table \ref{tab:tab1}). The odd rows in the figure are from NIHAO-UHD while even rows show the same galaxies in RiNG. Columns from the left to right represent the 2D density map of gas in edge-on view with white arrows for pixelized velocity vectors, the 2D temperature map averaged by gas density in edge-on view, the 2D metallicity map averaged by gas density in edge-on view, and the 2D zoom-in density map of gas disc in face-on and edge-on view, respectively. As shown in the most right panels, both codes consistently produce thin discs for most galaxies. All three cold discs in the NIHAO-UHD have well-developed flocculent arms that extend from a dense core, while the center of cold discs derived by \texttt{GIZMO} is relatively flat. From the face-on perspective, it can be seen that the gas density declines more rapidly at the edge of the NIHAO-UHD disc. The edges of disc derived by \texttt{GIZMO} appear relatively less distinct as they are surrounded by an ambient gas halo, the density of which is linked to the virial mass (we will present quantitative comparisons in Fig. \ref{fig:profile}). 

By examining the left three columns of edge-on maps, more intriguing features associated with the cold and hot gas in two simulations become apparent. Note that the dashed circles indicates the range of $0.15R_{\rm vir}$, which is adopted to divide the CGM and ISM environment (see definitions in Section \ref{sec:2.3}). Firstly, in NIHAO-UHD simulation, the hot and cold gas are distinctly spatially separated: the hot gas is distributed with many clumps, and the cold gas displays a filamentary structure. In contrast, the temperature map derived by \texttt{GIZMO} reveals a relatively smooth hot gas halo around the central disc. This hot gas halo is also clearly visible from the velocity vectors in the first column: the average velocity of gas particles located in the hot halos is smaller, indicating that the motion of hot halo is likely random, or in a ``kinetic hot state". Secondly, in NIHAO-UHD simulations, the spatial distribution of hot gas is closely linked to outflows in the velocity vectors map and high-metallicity regions, indicating that the hot gas in NIHAO-UHD is heated by recent nearby stellar feedback and chemical enriched by supernova remnants. We note that for the g2.79e12 system, the correlation among hot gas, outflows, and high-metallicity regions is not prominent compared to other two galaxies, suggesting that in this galaxy the hot gas is primarily maintained by gravitational heating.

%Specifically,  since they are actually the delayed cooling particles that are heated by recent nearby stellar feedback. Their cooling has not yet turned on. In contrast, the temperature map derived by GIZMO displays many rippled structures that illustrate the sharp shock front between hot feedback gas and cooled CGM gas. The most prominent feature in the third column is the large-scale outflows with high metallicity in the NIHAO simulations, while in GIZMO's metallicity maps the large-scale outflow can hardly be seen. 

\begin{figure*}[htbp]
\gridline{\fig{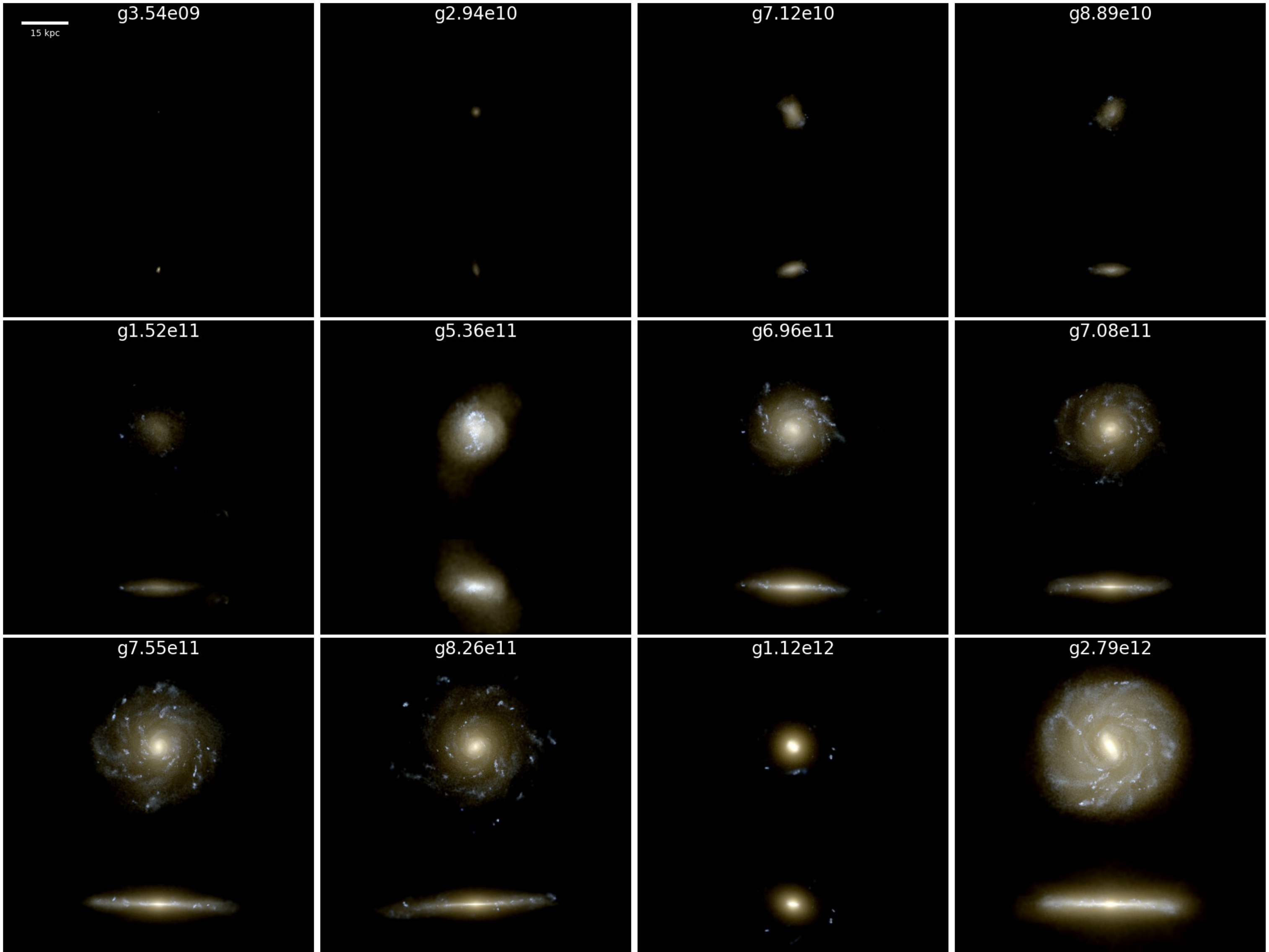}{0.75\textwidth}{(NIHAO-I and NIHAO-UHD)}}
\gridline{\fig{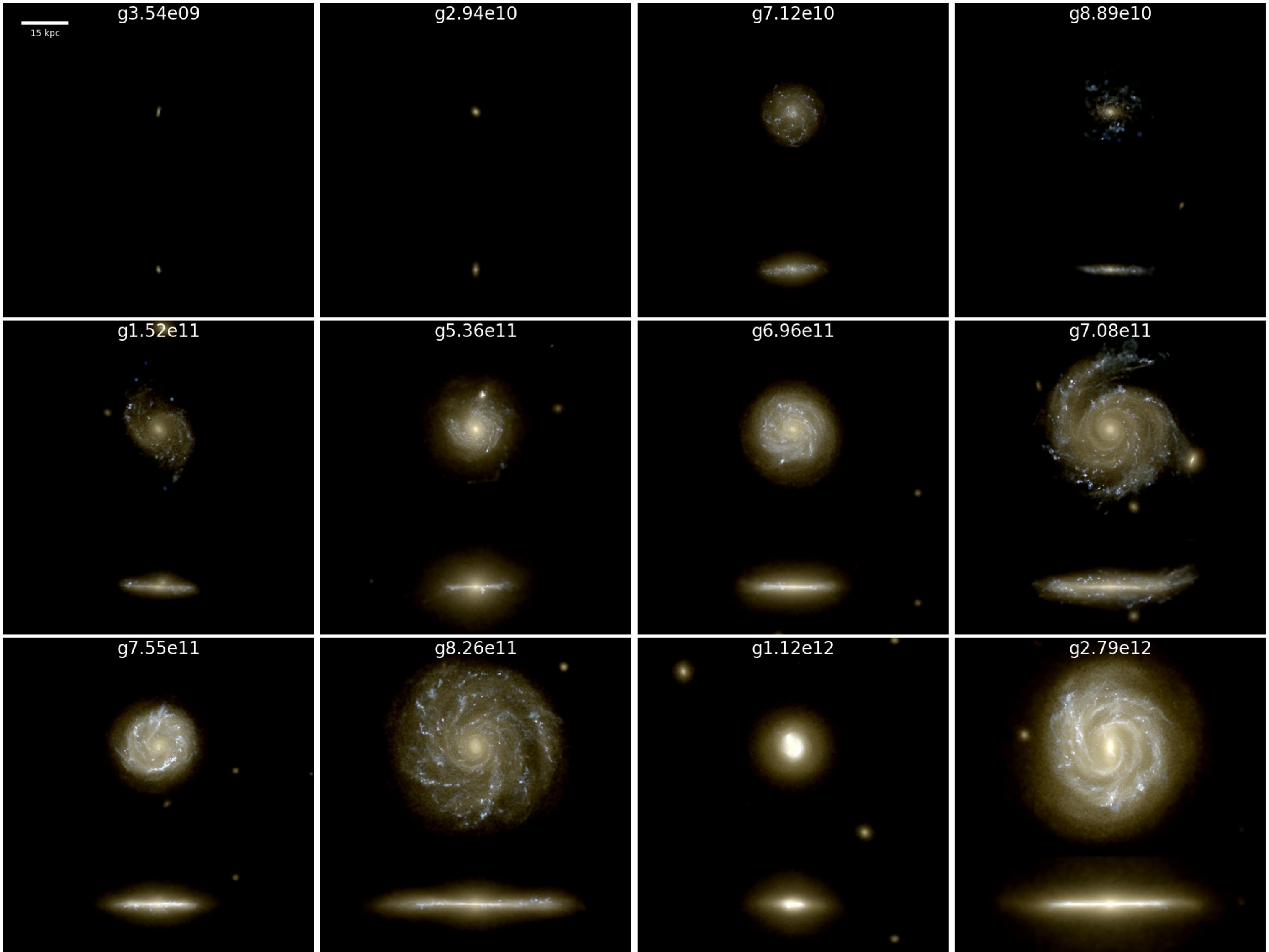}{0.75\textwidth}{(RiNG)}}
\caption{Comparison of mock optical images of all 12 systems of this work in the wavelength bands i, v, and u. Here we only plot the disc region with surface brightness lower than 26.5 magnitudes per square arc second.}
\label{fig:mocks}
\end{figure*}

The comparison of face-on and edge-on view of all 12 systems at $z=0$ are shown in Fig. \ref{fig:mocks}, where we show their mock optical images in the wavelength bands i, v, and u. It is found that disc galaxies are relatively more frequent in RiNG for halos with a virial mass of $10^{11\sim12}\mathrm M_{\odot}$, with the exception of g1.12e12. In fact, neither \texttt{GIZMO} nor \texttt{GASOLINE} produces a disc in this system, and it could be due to the lower spin parameter ($\sim 0.02$) for its host halo.

%the NIHAO simulation failed to produce discs in two systems (g5.36e11 and g7.12e10), and the disc size shows no clear trend with either halo mass or spin parameter. The reason for the failure of the g1.12e12 system to develop a cold disc in both simulations may be due to its too low spin parameter (0.020 in GIZMO and 0.024 in NIHAO).

\begin{figure}[htbp]
\plotone{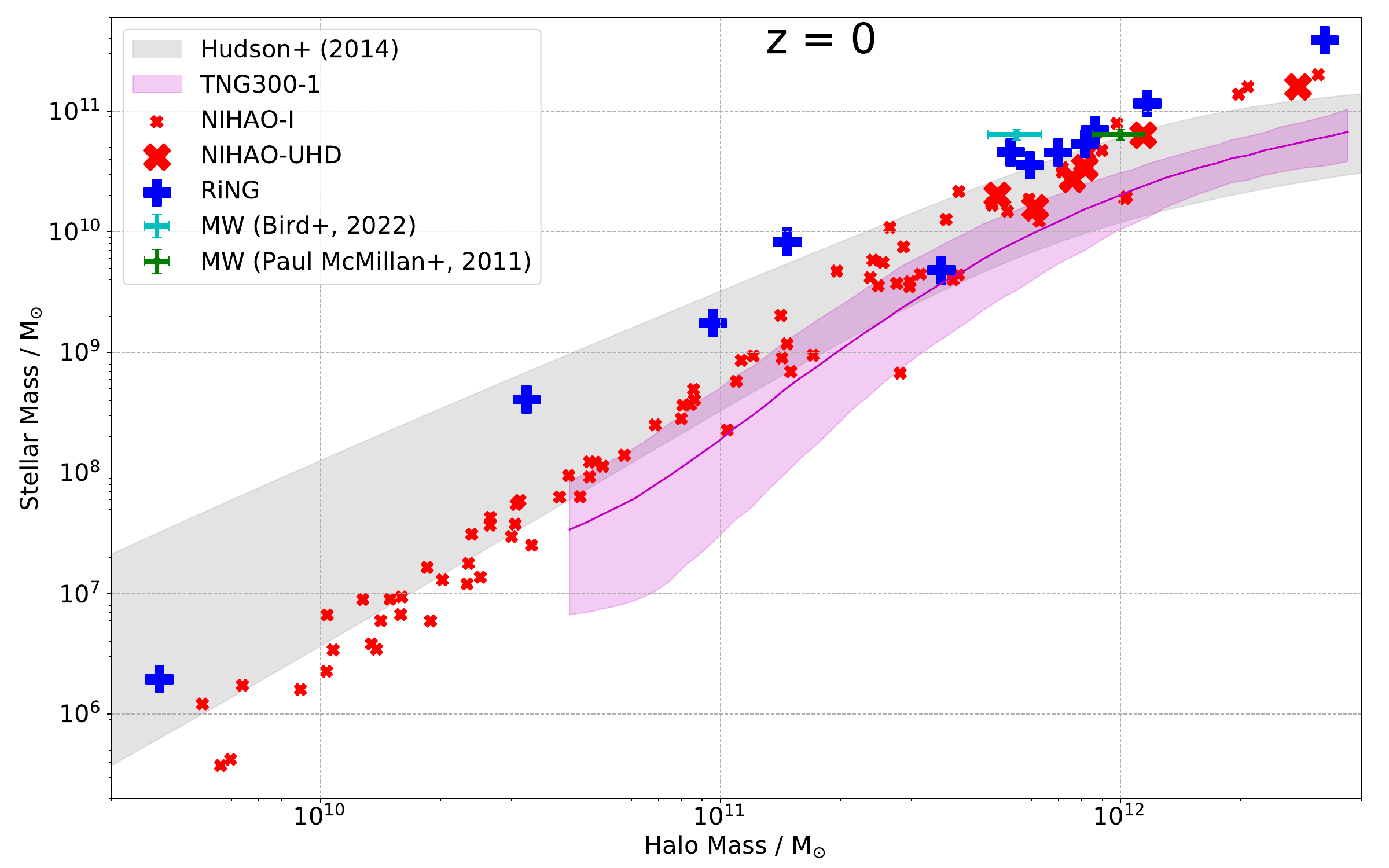}
\caption{Halo mass versus stellar mass relation for galaxies with halo mass larger than $10^{9.5}\,\mathrm{M_{\odot}}$. Shaded grey regions mark the observational results from the CFHTLenS galaxy lensing data \citep{2014ApJ...787L...5H}. As a comparison, results of all central galaxies from TNG300-1 simulation \citep{2018MNRAS.473.4077P}, as well as two recent observation results of halo mass and stellar mass of the MW are also shown. Within each halo mass bin, the upper and lower limits of TNG300-1 represent the 1st and 99th percentiles, respectively.}
\label{fig:hsm}
\end{figure}

The relation between halo mass and stellar mass for galaxies with  $M_{\rm vir}>10^{9.5}\,\mathrm{M}_{\odot}$ is shown in Fig. \ref{fig:hsm}. The results from RiNG and NIHAO suite are represented by blue plus signs and red crosses, respectively. For comparison, the observational results from the CFHTLenS \citep{2014ApJ...787L...5H} and the results from TNG300-1 simulation \citep{2018MNRAS.473.4077P}, as well as two recent observational results of the MW are also included. As we can see, the stellar masses obtained from \texttt{GIZMO} are systematically higher compared to TNG300-1, CFHTLenS and NIHAO suite. Considering the errorbars, the recent observations of stellar mass of MW from \cite{2022MNRAS.516..731B} are slightly greater than g5.36e11 and g6.96e11 galaxy in RiNG, and the results of \cite{2011MNRAS.414.2446M} are consistent with the NIHAO-UHD galaxy g1.12e12 and the RiNG galaxy g8.26e11. Overall, we do not see significant difference in the star formation efficiency between two simulations. 

\subsection{Phase Diagram of Gas} \label{sec:3.1}

\begin{figure*}[htbp]
\gridline{\fig{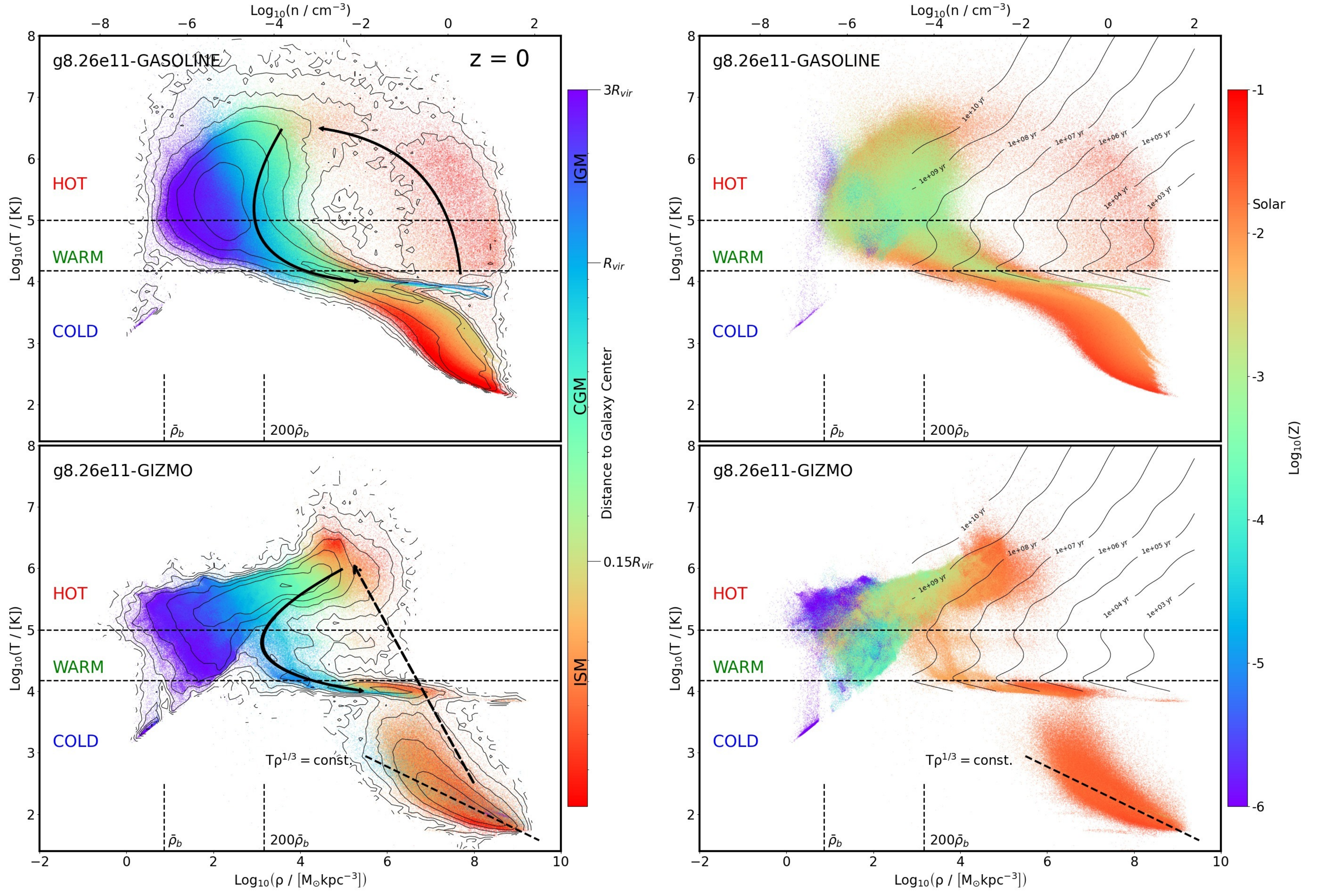}{1.0\textwidth}{}}
%\plotone{rt.pdf}
\caption{Comparison of density-temperature diagram of gas particles within $3R_{\rm vir}$ in g8.26e11 system. The regions for HOT, WARM and COLD are divided by horizontal dashed lines. The colors of the dots in the left and right panels denote the distance (in logarithm) to the galaxy center and metallicity (in mass fraction and also in logarithm), respectively. Thus the gas in ISM, CGM and IGM is approximately depicted in red, cyan, and blue hues in left panels. The left contour maps give the mass distribution of gas particles on the diagram, while the right contour maps show the cooling time of gas with solar metallicity under the assumption of collisional ionization equilibrium (CIE). The mean cosmic baryon density and its 200 times value are marked on the x-axis. The black dashed line in the lower panels demonstrates the best fitting polytropic state of the cold ISM.}
\label{fig:phase}
\end{figure*}

Previous section have shown that the morphology of MW-mass galaxies is marginally consistent between the two simulation, it is then interesting to compare the phase diagram of the gas, namely the distribution of density versus temperature. As an example, Fig. \ref{fig:phase} shows the comparison of phase diagram of galaxy g8.26e11 at $z=0$, including all gas particles within a distance of $3R_{\rm vir}$ to the galactic center. The colors of the dots in the left and right two panels denote the distance to the galactic center and metallicity (in mass fraction) of each gas particle, respectively. We also mark the mean cosmic baryon density $\bar{\rho}_b$ and $200\bar{\rho}_b$ on the x-axis, hence the dots on the right side of $200\bar{\rho}_b$ roughly illustrate the gas within the dark matter halo. It is seen that in both simulations of this system, metal poor gas can hardly be seen within $R_{\rm vir}$ (quantitatively, the mass fraction of metal poor gas with $Z<10^{-4}$ is less than $10^{-3}$ in all cases), while chemically enriched gas is scattered over all regions by accumulated stellar feedback, and even mixed with the metal poor IGM outside $R_{\rm vir}$. 

%Another prominent common feature concerns the primordial IGM gas, with some distributed along the adiabatic line $T\propto\rho^{2/3}$. As the primordial gas is approaching halos with $M_{\mathrm{vir}}\sim10^{11-12}\mathrm M_{\odot}$, two possible processes coexist, i.e. cold flow and shock heating \citep{2005MNRAS.363....2K}, which respectively corresponds to the dots in isotherm state around $10^4\,\mathrm{K}$, and the metal poor dots over the adiabatic line. Upon comparing the cold flows across all panels, the NIHAO cold flows are significantly more distant, denser, and also have lower metallicity compared to those produced by GIZMO. 

It is seen that the main difference between the two simulations lies in the ISM (depicted in red hues in left panels) and the hot CGM (cyan hues in hot region). Firstly, due to different star formation density threshold, fine structure cooling and molecular cooling functions, the cold ISM in RiNG experienced an effective polytropic process with index $\gamma\simeq4/3$, and finally reaches a denser and colder state at the lower right corner of the diagram. The best slope of cold ISM in RiNG galaxy g8.26e11 is fitted by gas particles that satisfy $T<10^3\,\mathrm{K}$ and $\rho>10^6\,\mathrm{M}_{\odot}\mathrm{kpc}^{-3}$, which is in accordance with the default effective sub-grid state of multi-phase ISM used in the EAGLE project \citep{2008MNRAS.383.1210S, 2015MNRAS.446..521S}. In NIHAO simulations, the cold ISM gas is in an isobar state (see Fig. 4b in \cite{2019MNRAS.485.2511T}) before reaching the density threshold of star formation. 

It is reasonable to expect that the cold ISM, influenced by recent neighboring stellar feedback in two simulations, would undergo chemical enrichment and heating, thereby transitioning into the hot ISM/CGM. In Fig. \ref{fig:phase} we use black arrows to designate a typical evolution path of feedback gas in two simulations. In the NIHAO-UHD case, the majority of heated gas particles initially exist in a hot yet dense state (referred to as the warm/hot ISM), then move to low-density regions through the mechanical work accomplished by the pressure difference (referred to as $p\mathrm{d}V$ work hereafter), leading to outflows (as discussed in the next section), before eventually returning to an isotherm state around $10^4\,\mathrm{K}$ by cooling. While in the RiNG case, the temperature of hot ISM is higher than $10^6\,\mathrm{K}$ with density lower than the NIHAO-UHD case. In the lower left panel we use a straight dashed line to represent the direct transition of heated gas from the cold ISM to the hot region by avoiding the warm phases. Following this transition, the gas will then accumulate in the hot CGM (also refer to the next section) before its final cooling and recycling back to the isotherm state around $10^4\,\mathrm{K}$.

For ideal gas with adiabatic index $\gamma=5/3$, temperature $T$ (unit in $\rm K$) and number density $n$ (unit in $\rm cm^{-3}$), the radiative cooling time formula is given by:
\begin{equation}
    t_{\mathrm{cool}}\simeq\frac{3.3\;T}{n\,\Lambda_{-23}(T)}\;\mathrm{yr},
    \label{equ1}
\end{equation}
where $\Lambda_{-23}$ is the cooling rate (unit in $10^{-23}\,\mathrm{erg}\,\mathrm{cm}^3\mathrm{s}^{-1}$). From the upper panel of Fig. \ref{fig:phase}, the typical thermal state of recently heated hot ISM has $T=10^{4\sim6}\,\mathrm{K}$, $n=10^{-1\sim1}\,\rm cm^{-3}$, and for which the cooling rate $\Lambda_{-23}$ is about $10^{0.5\sim1.6}$ with an assumption that $Z=Z_{\odot}$. So the cooling time of recently heated gas in NIHAO-UHD is expected to be $10^{3\sim6}$ years (for detailed values please refer to the contour maps in right panels of Fig. \ref{fig:phase}), it indicates that the thermal energy from stellar feedback would be radiated away in a couple of time steps if the cooling is not delayed. As a result, the clustering of heated gas in the upper-right region of the NIHAO-UHD diagram is an artifact of the delayed cooling model \citep{2006MNRAS.373.1074S}. In contrast, RiNG predicts cooling times for hot ISM gas between $10^{6\sim9}$ years (as seen in the lower-right panel). This extended cooling timescale allows feedback gas to maintain its thermal energy for a considerable period. Due to the significantly higher pressure of heated gas compared to the cold ISM, it cannot stably accumulate in dense regions. Ultimately, in both simulations, the heated gas evolves towards less dense regions: transitioning into the hot CGM.

\subsection{Hot CGM and Inflow/Outflow Gas} \label{sec:3.2}

\begin{figure}[htbp]
%\gridline{\fig{g8.26e11.pdf}{0.5\textwidth}{}}
\plotone{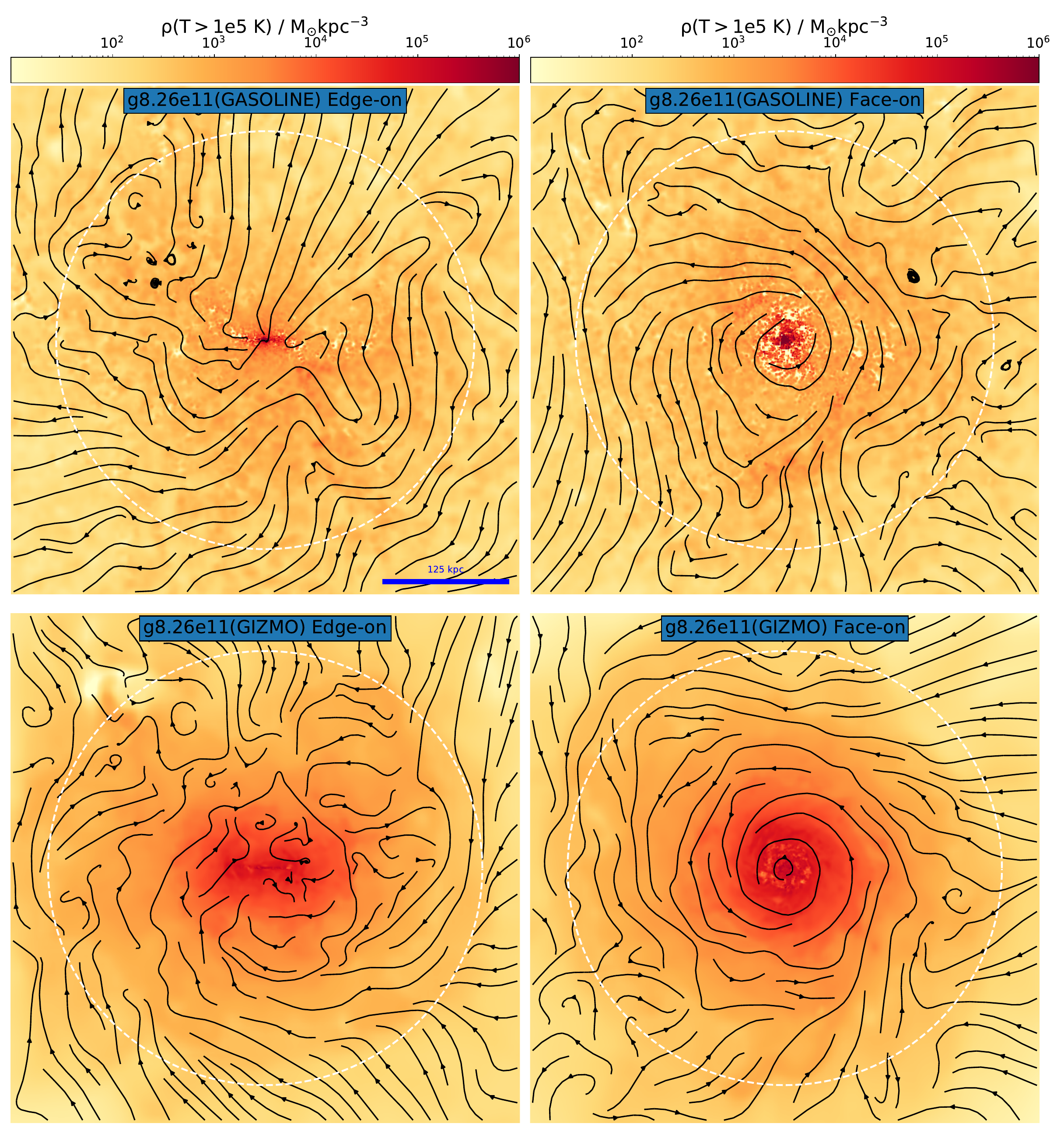}
\caption{Comparison of the hot CGM distribution in g8.26e11. Colors represent the volume density of gas with temperature higher than $10^5\,\mathrm K$. The streams in all four panels represent the flow distribution. The left two panels are in edge-on view, and the right two panels are in face-on view. The white dashed circle designates the halo virial radius.}
\label{fig:hothalo}
\end{figure}

To investigate the whereabouts of the feedback gas in the subsequent evolution, in Fig. \ref{fig:hothalo} we first compare the volume density of hot gas ($T>10^5\,\mathrm K$). As mentioned in previous section, the hot gas in NIHAO-UHD predominantly clusters around the central disc (i.e. hot ISM). Some of the hot gas particles experience acceleration through $p\mathrm{d}V$ work before being expelled from the dark matter halo along the vertical direction (perpendicular to the disc plane) and those process subsequently reduces the amount of CGM gas (see Fig. \ref{fig:ma} for baryon fraction). While for the corresponding galaxies derived by \texttt{GIZMO}, as demonstrated in the lower two panels of Fig. \ref{fig:hothalo}, a net inflow at the virial radius interacts with the feedback gas internally, which finally results in an inner CGM gas halo with relatively smooth distribution. A similar conclusion is also mentioned by \cite{2023arXiv230108263H}.

\begin{figure}[htbp]
\plotone{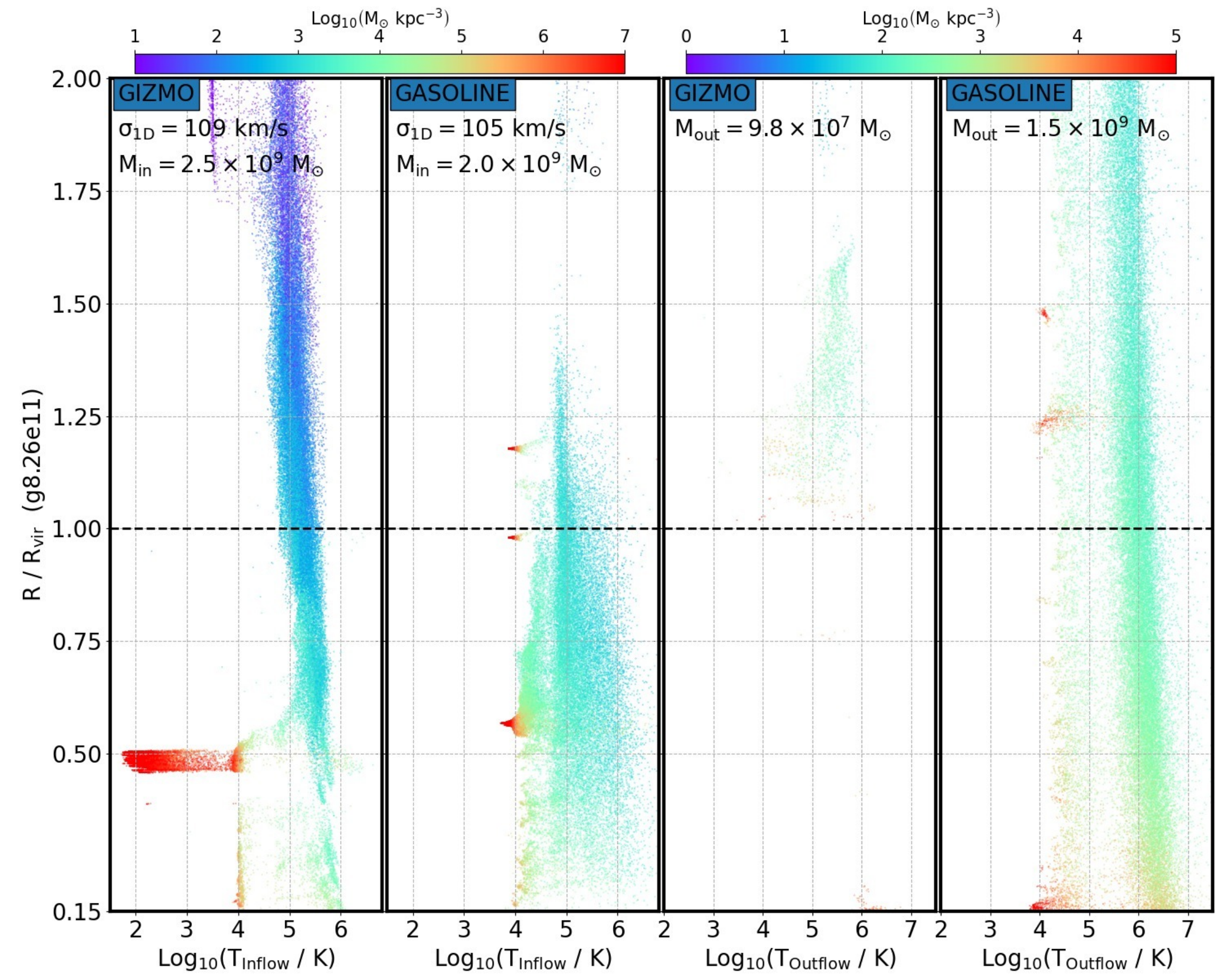}
\caption{The temperature distribution of the inflow and outflow gas at different radii is depicted with varying colors indicating the density of gas particle. The 1D velocity dispersion of the halo, as well as the mass of total inflow/outflow gas within $2R_{\rm vir}$ is denoted in each panel.}
\label{fig:inflow}
\end{figure}

On the other hand, the CGM environment plays a crucial role in shaping both inflow and outflow gas. To identify the inflow and outflow gas, we strictly follow the methodology outlined in \cite{2015MNRAS.454.2691M}. Specifically, we classify gas particles as inflow or outflow based on the following inequalities:
\begin{equation}
v_{\mathrm{rad}}\equiv\boldsymbol{v\cdot}\frac{\boldsymbol{r}}{r}\left\{\begin{aligned}>&\,\sigma_{\mathrm{1D}},\quad &\mathrm{outflow}\\
<&-\sigma_{\mathrm{1D}},\quad &\mathrm{inflow}\end{aligned}\right.
\end{equation}
where $\boldsymbol{v}$ and $\boldsymbol{r}$ are velocity and position of the gas particle relative to the galactic center. $\sigma_{\rm 1D}$ is the 1D velocity dispersion of the halo. 

Fig. \ref{fig:inflow} depicts the distribution of inflow and outflow gas at various radii at $z\sim 0$ for the same galaxy, g8.26e11. The color of the dots are correlated with the density of the gas (as shown in the horizontal color bar). Note that there are some dots clustered around a particular radius and distributed along the horizontal axis, which are gas particles from satellite galaxies. We do not exclude them from the plot but here we focus on the smooth component of the inflow/outflow gas which is mainly distributed vertically.

%are from satellite galaxies, and the smooth. The varying colors represent the density distribution of inflow and outflow particles. In this visualization, the inflow/outflow gas appear as vertically extended structures, while the gas in infalling/departing satellite galaxies are represented horizontally. 
Notably, as shown by the label in each panel, both simulations exhibit a similar amount of inflow gas of $\sim [2-2.5] \times 10^{9}M_{\odot}$ within $2R_{\rm vir}$, while  the amount of outflow mass shows a substantial discrepancy, with near one order of magnitude for this galaxy, as seen in the right two panels. For the inflow gas, the comparison show major difference that for RiNG, a considerable amount of inflow gas exists at a distance as far as $2R_{\rm vir}$, whereas in NIHAO-UHD, there is few inflow gas beyond $1.5R_{\rm vir}$. As shown in the work by \cite{2016MNRAS.456.3542T}, this suppression of inflow in NIHAO-UHD is due to the presence of strong outflows (as seen in the right panel), which prevent the IGM gas from falling efficiently. Therefore, most of the inflow gas within $R_{\rm vir}$ in NIHAO-UHD simulation is the recycling gas that has been heated by previous stellar feedback.
%The lower left panel in Fig. \ref{fig:sfhr} reveals that galaxy g8.26e11 exhibits a star formation rate (SFR) of $1\sim2\,\mathrm{M_{\odot}yr^{-1}}$. This demonstrates that even in a MW-like quiescent disc galaxy simulated by NIHAO, stable outflows can occur. Furthermore, the net gas accretion rate from the cosmic web is roughly equivalent to the difference between inflow and outflow rates. Consequently, the sustained outflows driven by NIHAO simulations effectively inhibit significant net gas accretion from the cosmic web. 

The temperature distribution of inflow/outflow gas is also different in the two simulations. The left panel shows that in RiNG simulation, the infalling gas gradually heats up from $10^5\,\rm K$ to nearly $10^6\,\rm K$, aligning with the temperature of the hot gas halo mentioned previously. Conversely, NIHAO-UHD simulation finds that the majority of infall gas is cooling from $10^5\,\rm K$ to $10^4\,\rm K$, with only a negligible fraction experiencing heating.

\subsection{Galaxy Density Profiles} \label{sec:3.3}

\begin{figure}[htbp]
\plotone{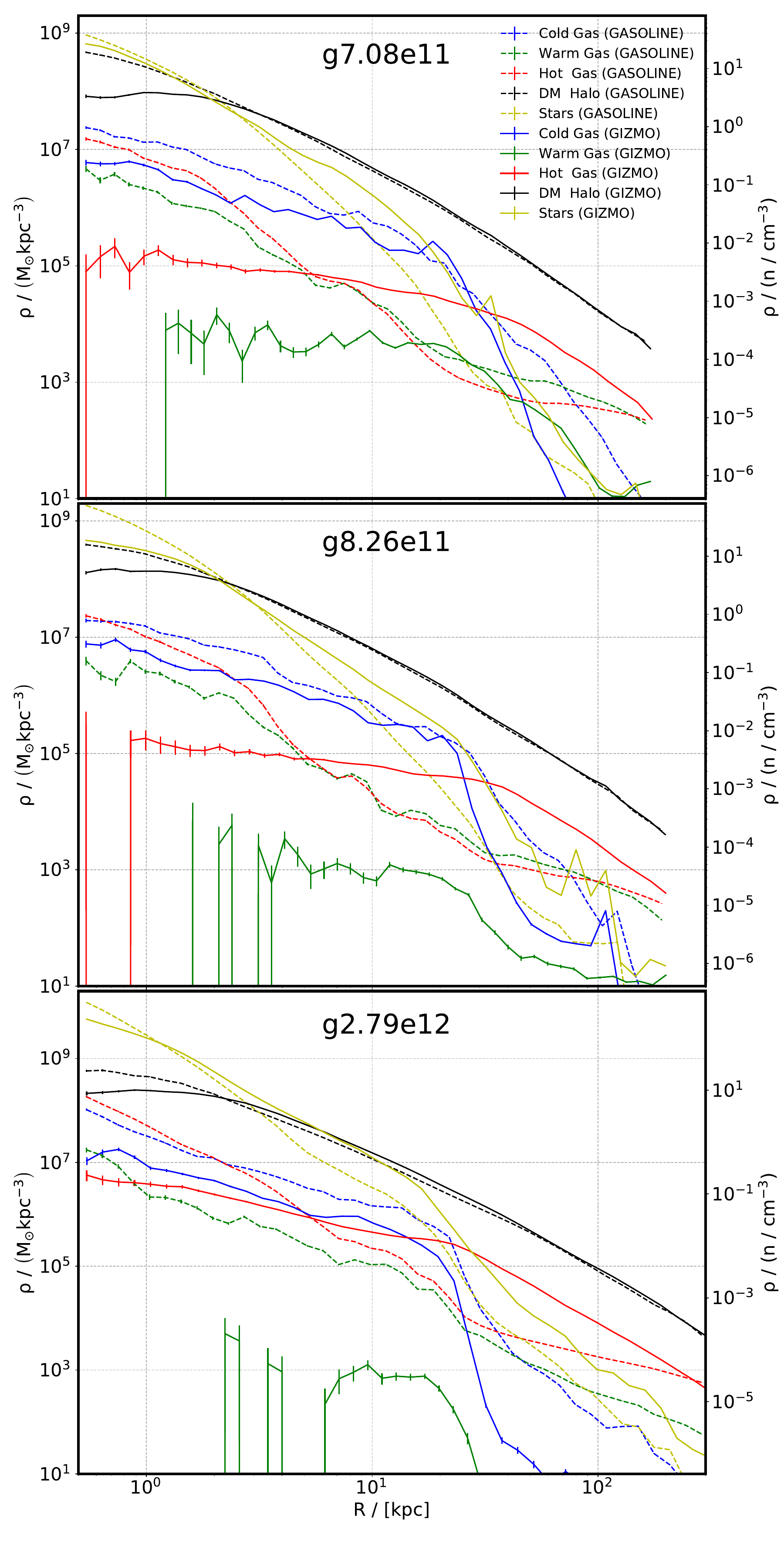}
\caption{The 3D density profile of star and gas with different temperatures is compared across two simulations for three representative systems (g7.08e11, g8.26e11, g2.79e12). The dark matter halo profile is shown in each panel for comparison with the profile of the hot gas halos. Solid lines and dashed lines denote samples run by \texttt{GIZMO} and \texttt{GASOLINE}, respectively.}
\label{fig:profile}
\end{figure}

The 3D profile of star, dark matter, as well as gas with different temperatures (i.e. cold, warm and hot) is compared in Fig. \ref{fig:profile} for three systems. The components that differ the most between the two simulations are warm and hot gas. In NIHAO-UHD simulation，the profiles of hot and warm gas exhibit a consistent trend from inner to outer regions. Note that the turning point of the decreasing profile of cold gas roughly quantifies the extent of the cold disc, within which there is a significant amount of hot and warm gas concentrated in the disc region. But for the RiNG samples, there is significantly less warm gas for all the three galaxies. The profile of hot gas roughly exhibits a double power-law distribution: relatively flat in the interior and parallel to the dark matter profile outside the cold disc. 
%This result is in agreement with the hot gas halo presented in Fig. \ref{fig:hothalo}. 

\begin{figure}[htbp]
\gridline{\fig{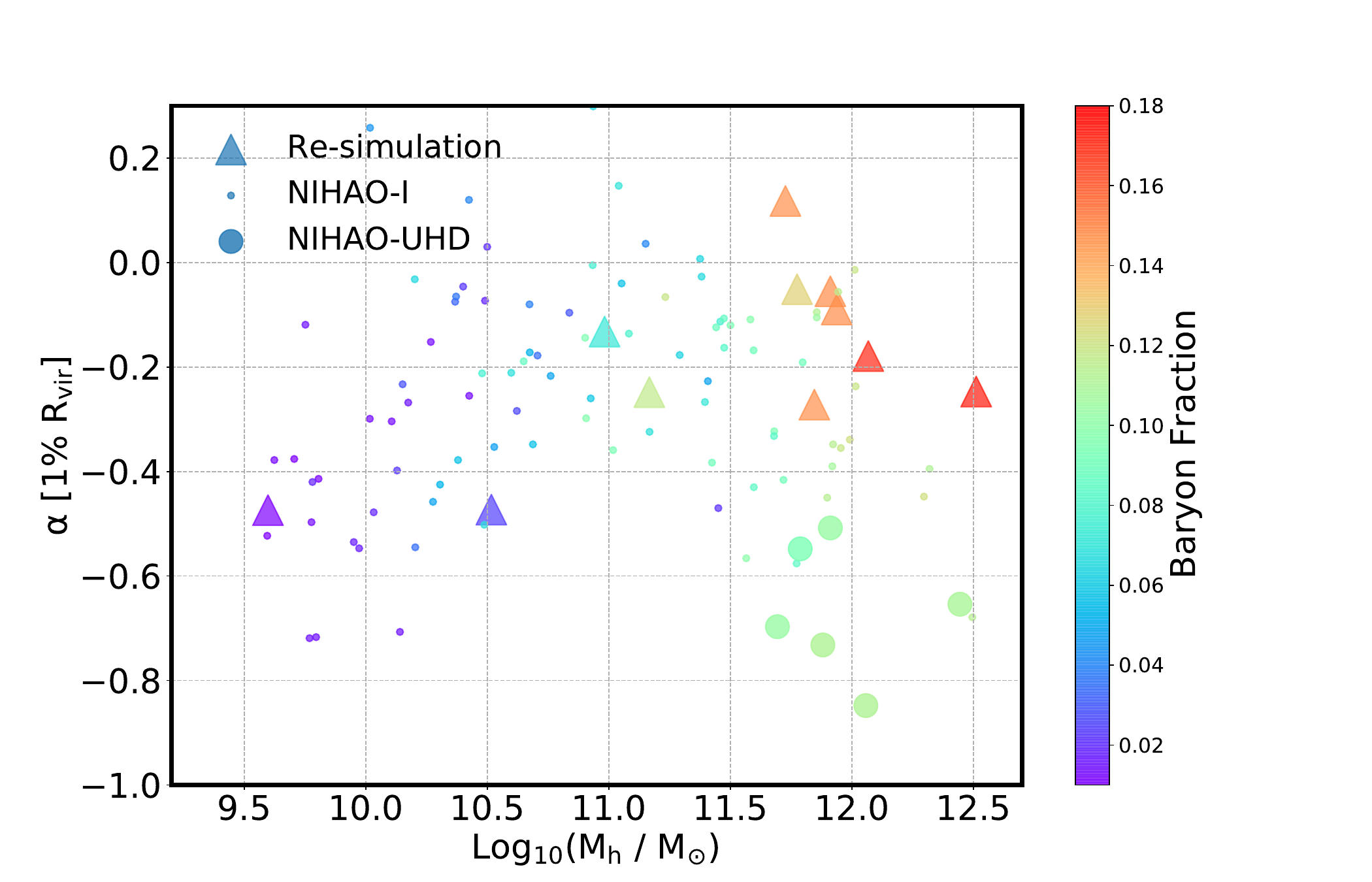}{0.55\textwidth}{}}
%\plotone{ma.pdf}
\caption{The impact of stellar feedback on the inner halo dark density profile and baryon mass fraction. The y-axis gives the best fitting value of the inner dark matter slope $\alpha$ within $0.01\,R_{\mathrm{vir}}$, and the x-axis is the corresponding virial mass at $z=0$. Dots are galaxies in NIHAO-I and NIHAO-UHD simulation and triangles are the corresponding galaxies run with \texttt{GIZMO}. The color is correlated with the baryon fraction within $R_{\mathrm{vir}}$, as labelled in the right vertical color bar.}
\label{fig:ma}
\end{figure}

The outflows triggered by stellar feedback have the ability to disturb the central gravitational potential and transfer energy to dark matter particles, thereby altering the inner profile of the halo \citep{1996MNRAS.283L..72N, 2012ApJ...744L...9M}. This process has been well shown to result in a core-like profile within the dark matter halo \citep{2012MNRAS.422.1231G, 2014MNRAS.437..415D, 2014ApJ...786...87B, 2016MNRAS.456.3542T, 2017MNRAS.471.3547F}. In Fig. \ref{fig:ma} we further show the slope of the dark matter halo at the radius of $0.01R_{\rm vir}$ from our galaxies, other samples from NIHAO suite are also added for comparison. The dots are for results of NIHAO-I and NIHAO-UHD, and triangles are for the corresponding galaxies in RiNG. It is seen that both two simulations could produce cores in these galaxies, and their difference on the inner slope is very minor for halo with mass  $M_{\rm halo} < 10^{12}M_{\odot}$. For galaxies with $M_{\rm halo}\sim 10^{12}M_{\odot}$, \texttt{GIZMO} could produce more flatter profile. This is in agreement with the results from the FIRE-2 simulation that a flat core is found in the inner $\sim 2\rm kpc$ of MW size halo \citep{2020MNRAS.497.2393L}.

\subsection{Recycle of Feedback Gas } \label{sec:3.4}

\begin{figure*}[htbp]
\gridline{\fig{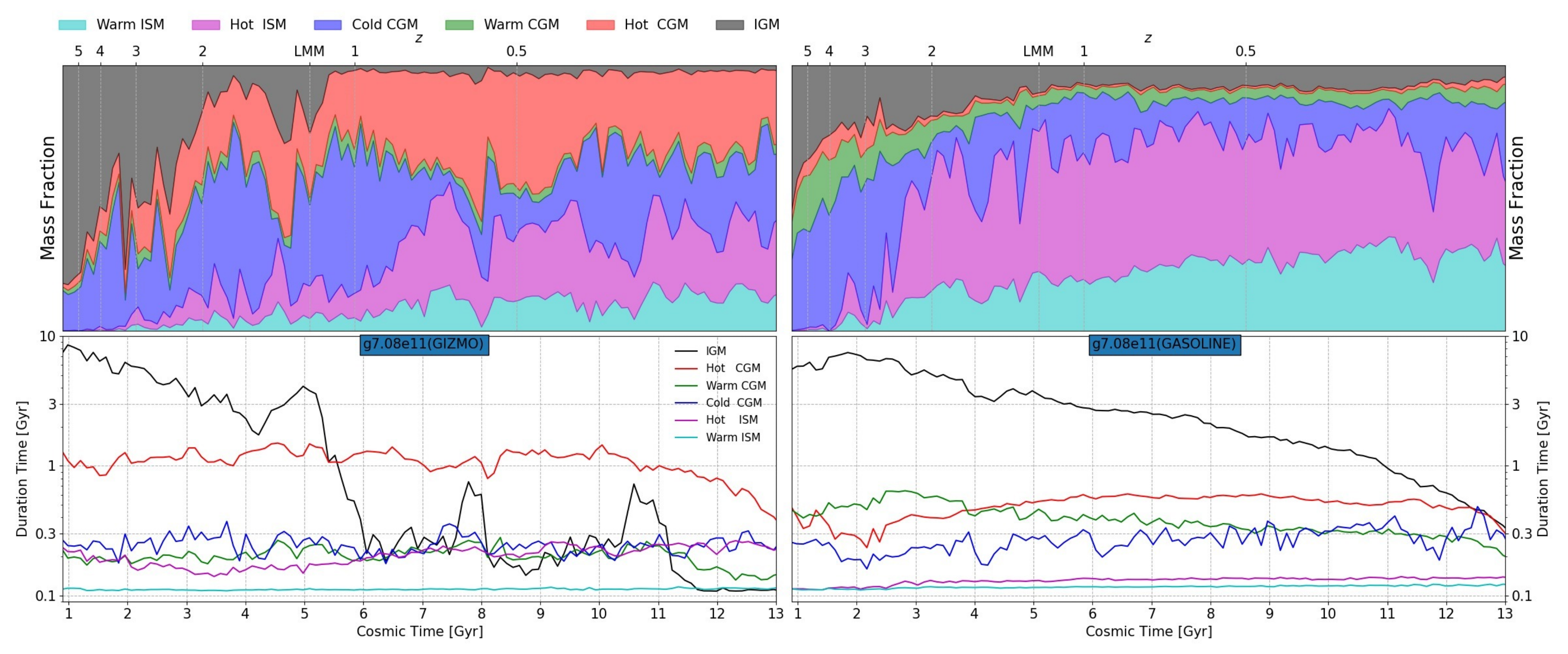}{1.0\textwidth}{(a. g7.08e11)}}
\gridline{\fig{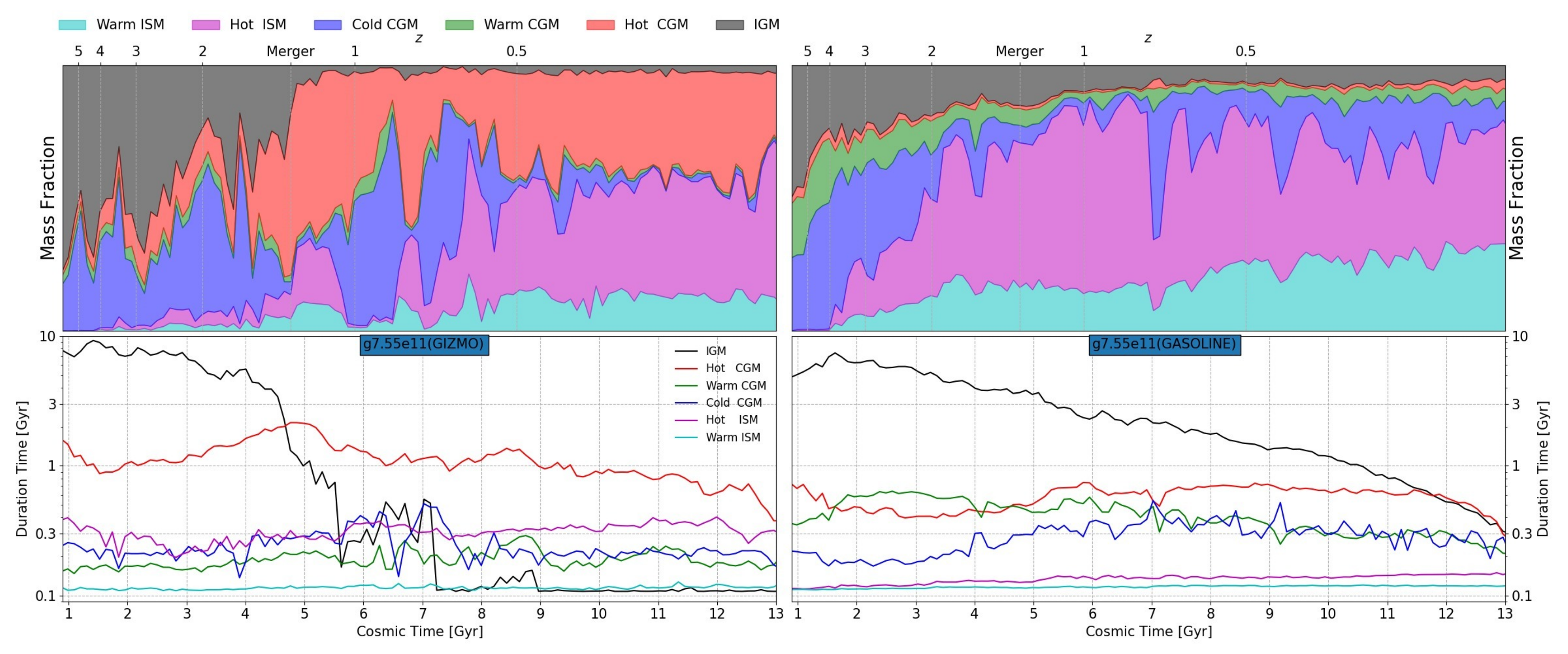}{1.0\textwidth}{(b. g7.55e11)}}
\caption{Comparison of feedback gas in different phase states. Subplots (a) and (b) are featuring two galaxies, g7.08e11 and g7.55e11, run by \texttt{GIZMO} and \texttt{GASOLINE} respectively. For each galaxy, the upper panel is the mass fraction of feedback gas in different outermost phase states (see text for the meaning of outermost phase state), as labelled in each panel, and lower panel is the duration of feedback gas in each phase.}
\label{fig:feedback}
\end{figure*}

In hydro-dynamical simulations, the feedback from evolved stars, such as stellar wind and supernova explosion, will release both mass and energy to the nearby cold ISM. As those feedback processes happen on scales usually far below the spatial resolution of simulations, different assumptions on the sub-grid processes are often adopted in different models or hydro-dynamical codes, which have been shortly mentioned in Section \ref{sec:1}. Therefore, the cold gas in the ISM around star forming region will acquire different energy or momentum, and resulting in quite different subsequent evolution in different models. Depending on the feedback model and how much energy gained, the feedback affected cold gas, hereafter called as feedback gas, will have different thermal states and velocities in different hydro-dynamical simulations. For example, the feedback gas could be heated to high temperature and is identified as hot ISM if its position remains within the star forming region. However, if the feedback gas gain high velocity, it could escape from the star forming region and being identified as either hot CGM, warm CGM, cold CGM or even IGM, depending on how far it can travel from galactic center. In this part, we focus on the different fates of the feedback gas in two simulation suites.

Using the standard methods of tracing particles across snapshots in Lagrangian schemes via their unique particle ID, we identify the feedback gas  (as mentioned in Section \ref{sec:2.3}) in each snapshot and trace its subsequent evolution. To quantify the degree of gas affected by the feedback processes, we rank the gas phases that feedback gas is likely to reach based on its temperature and distance to galactic center. Overall, the feedback gas could reach these following phase status depending on how strong the feedback process is, that is warm ISM, hot ISM, cold CGM, warm CGM, hot CGM and IGM. In general, feedback gas is more easy to reach the warm ISM status, but being difficult to reach the IGM status unless the feedback is very strong and the feedback gas gain very high velocity. For a specific snapshot, we trace the evolutionary trajectories in various phase states for all feedback gas (identified based on the criteria outlined in Section \ref{sec:2.3}) until it transitions back to cold ISM or forms stars, or until we reach the last step of the simulation. Moreover, we seek the outermost phase that the feedback gas can attain during the evolutionary trajectory and categorize the feedback gas accordingly.

In Fig. \ref{fig:feedback} we show the comparison of feedback gas in different phase states, featuring two MW-mass galaxies from two simulations with the g7.08e11 in subplot (a) and g7.55e11 in subplot (b). Each subplot has two panels: the mass fraction of feedback gas in different outermost phase states, and the duration in each phase state. Although the results for the two galaxies has many differences in details, we use g7.08e11 galaxy as an example to illustrate the common features. It is seen from the upper panel in subplot (a) that at high redshifts ($z>3$), more than half of the feedback gas escapes the dark matter halo and is identified as the IGM. In the early stages, when the galaxy's gravitational potential was shallow, feedback-induced outflows could escape the galaxy more easily. This phenomenon is consistently predicted by both NIHAO-UHD and RiNG. However, after $z=3$, it is found that the feedback gas was in different states in different simulations. For example, \texttt{GIZMO} predicts a considerable fraction of feedback gas is in hot CGM phase, while this component is negligible in NIHAO-UHD sample. Besides, the fraction of feedback gas in IGM predicted by \texttt{GASOLINE} is higher compared to \texttt{GIZMO}. 

The lower panel of subplot (a) shows the duration of feedback gas in different phase state (not the outermost phase). The most noticeable difference between the two simulations lies in the duration for IGM (black line) and hot CGM (red line). In the RiNG simulation, although the feedback gas can stay in the IGM for several Gyr at high redshift, it exhibits a rapid decline to less than $1\rm Gyr$ after a certain redshift. In contrast, in NIHAO-UHD simulation, feedback gas can still remain in the IGM for over $1\rm\, Gyr$. This discrepancy arises from the different feedback mechanisms implemented in two simulations. In the mechanical feedback model of \texttt{GIZMO}, the supernova ejecta has fixed velocities (only depending on their type, refer to Appendix A of the FIRE-2 paper, i.e., \cite{2018MNRAS.480..800H}). As the gravitational well of the host galaxy deepens with cosmic time (or decreasing redshift), feedback gas in IGM will be more easily to be accreted by the host galaxy. For the galaxy g7.08e11, it is found that the abrupt decrease of the duration is caused by a major merger of the host galaxy which leads to a sudden deepening of the gravitational well. In contrast, in NIHAO suite, the final velocity of the supernova ejecta is determined by the total $p\mathrm{d}V$ work during the period of delayed cooling, which is less sensitive to the sudden deepening of the gravitational well, but show a smoothing decreasing with time.

%It should be noted that for MW-mass systems without LMM (i.e., of which mass accretion is primarily through minor mergers and smooth accretion), the relationship between merger events and the transition of feedback gas in IGM may not as prominent as shown here.

The duration of feedback gas confined within CGM also shows difference between the two simulations. For MW-mass galaxies in RiNG simulation, feedback gas can persist in hot CGM for approximately 1Gyr, which is about double the time in NIHAO-UHD. Additionally, while the duration of warm feedback gas in the ISM is quite short in both simulations, the NIHAO-UHD shows a wider range for the duration of warm gas in the CGM. In contrast, \texttt{GIZMO} predicts about $0.2\rm Gyr$ for this duration. This difference partly explains the scarcity of warm gas in the RiNG samples (see Fig. \ref{fig:profile}). Lastly, the duration of feedback gas in the cold CGM is roughly consistent between the two simulations.

\begin{table*}[htbp]
\caption{Mean duration (in Gyr) of feedback gas in different phases since $z=3$. The last column gives the averaged duration of feedback gas in all phases, so it is equivalent to the recycle time for ejected gas.}
\label{tab:tab2}
\centering
\begin{tabular}{cc c c c c c c c}
\hline\hline
\multicolumn{2}{c}{system/simulation} & warm ISM & hot ISM & cold CGM & warm CGM & hot CGM & IGM & averaged \\ \hline
\multicolumn{1}{c|}{g6.96e11} & NIHAO-UHD & 0.12 & 0.13 & 0.29 & 0.42 & 0.58 & 2.40 & 0.96   \\  
\multicolumn{1}{c|}{ }  & RiNG & 0.11 & 0.26 & 0.26 & 0.19 & 1.30 & 1.57 & 1.45 \\ \hline
\multicolumn{1}{c|}{g7.08e11} & NIHAO-UHD & 0.12 & 0.13 & 0.27 & 0.37 & 0.49 & 2.43 & 0.80   \\
\multicolumn{1}{c|}{ }  & RiNG & 0.11 & 0.21 & 0.25 & 0.20 & 1.06 & 1.19 & 1.27 \\ \hline
\multicolumn{1}{c|}{g7.55e11} & NIHAO-UHD & 0.12 & 0.14 & 0.29 & 0.40 & 0.56 & 2.27 & 0.81   \\ 
\multicolumn{1}{c|}{ }  & RiNG & 0.12 & 0.31 & 0.24 & 0.19 & 1.10 & 1.48 & 1.92  \\ \hline
\multicolumn{1}{c|}{g8.26e11} & NIHAO-UHD & 0.32 & 0.14 & 0.31 & 0.37 & 0.66 & 1.69 & 0.56   \\
\multicolumn{1}{c|}{ }  & RiNG & 0.11 & 0.25 & 0.42 & 0.14 & 0.94 & 0.43 & 0.92 \\ \hline
\multicolumn{1}{c|}{g2.79e12} & NIHAO-UHD & 0.13 & 0.19 & 0.29 & 0.50 & 2.63 & 2.02 & 0.60   \\ 
\multicolumn{1}{c|}{ }  & RiNG & 0.11 & 0.51 & 0.19 & 0.15 & 0.88 & 0.55 & 1.51  \\ \hline\hline
\end{tabular}
\end{table*}

As a further comparison, in Table \ref{tab:tab2} we list the mean duration of feedback gas in five MW-mass galaxies (g6.96e11, g7.08e11, g7.55e11, g8.26e11 and 2.79e12). The values here are averaged by feedback gas that stays in each phase since $z=3$. The last column is the averaged duration of all feedback gas regardless of the phase it stays, so that is the recycle time for feedback gas to be available for further star formation. Overall, our conclusions from the g7.08e11 in Fig. \ref{fig:feedback} also apply to other galaxies, that is the duration of feedback gas in IGM is longer in NIHAO-UHD than in RiNG while the duration of hot CGM is lower in NIHAO-UHD than in RiNG. The average recycle time of feedback gas (last column) predicted by \texttt{GIZMO} is about $50\% -150\%$ longer. The only exception involves the much longer duration of the hot CGM in g2.79e12 galaxy, which notably surpasses that of other NIHAO-UHD galaxies. Recalling the corresponding panels in Fig. \ref{fig:fig1}, the hot gas in this system is primarily maintained by gravitational heating rather than feedback gas. As a result, our previous discussions on feedback gas in hot phase are inapplicable to this particular system.

Combing results shown in Fig. \ref{fig:feedback} and in other previous figures, we can form a comprehensive understanding of how baryon recycles in MW-mass galaxies in the two simulations. Firstly, about the formation of hot gas halo. From Fig. \ref{fig:phase}, Fig. \ref{fig:inflow} and Fig. \ref{fig:profile}, it is found that the thermal state of the inner region of the hot gas halo in RiNG is to be approximately $T\sim10^6\,\rm K$ and $n\sim10^{-3}\rm cm^{-3}$. Therefore, the cooling time of hot CGM should be around $t_{\rm cool}\simeq0.6\,\rm Gyr$ (assuming a mean metallicity of $0.3Z_{\odot}$). It is shorter than what we have found from \texttt{GIZMO} predictions (red line in left panels of Fig. \ref{fig:feedback} or Table \ref{tab:tab2}), which indicates that the hot gas halo must get energy supply, a.k.a the stellar feedback. Besides, if the radius of the hot gas halo is $60\rm kpc$ and the gas particles move at approximately the virial velocity, then the crossing time of the hot halo is expected to be less than $0.3\,\rm Gyr$, also significantly shorter than the duration of feedback gas in this phase. This further reinforces the conclusion drawn from Fig. \ref{fig:fig1}: for RiNG galaxies, the movement of gas particles in the hot gas halo must be random and in a ``kinetic hot state".

Secondly, it is noteworthy that after $z\sim3$, the total mass of feedback gas in NIHAO-UHD is typically several times larger than that in RiNG. Most of this feedback gas remains confined within $0.15R_{\rm vir}$, with recycling periods not exceeding $0.2\rm Gyr$. A small fraction of feedback gas attains very high velocities and is vented beyond the virial radius as outflows. These feedback gases recycle on a significantly larger scale, effectively hindering substantial gas accretion from the cosmic web. The recycling period of this feedback gas is an order of magnitude longer than the previous recycle. For galaxies simulated by \texttt{GIZMO}, nearly all feedback gas remains confined within $R_{\rm vir}$ after a certain redshift, usually the last major merger or other previous mergers. The hot gas halo forms through a mixture of inflows, the infall of previously feedback gas and the recently feedback gas, with recycling periods extending up to $1\rm Gyr$. 

Finally, the significantly larger mass of feedback gas in NIHAO-UHD suggests that the outflow escaping the virial radius is non-negligible (as evidenced by the low but stable mass fraction of IGM in NIHAO-UHD galaxies, as well as the strong outflow depicted in Fig. \ref{fig:inflow}). Taking g7.08e11 as an example, by simply adding the mass of escaped feedback gas in each snapshot, about  $2.3\times10^{10}\,\mathrm{M}_{\odot}$ and $3.3\times10^9\,\mathrm{M}_{\odot}$ baryon have escaped the halo in the NIHAO-UHD and RiNG simulation, respectively. This leads to a decrease of total baryon fraction by $3.3\%$ and $0.46\%$ in two simulations, which partly explains the systematic differences of baryon fraction of MW-mass samples, as mentioned in Section \ref{sec:3.3}.

\subsection{Star Formation History} \label{sec:3.5}

\begin{figure*}[htbp]
\plotone{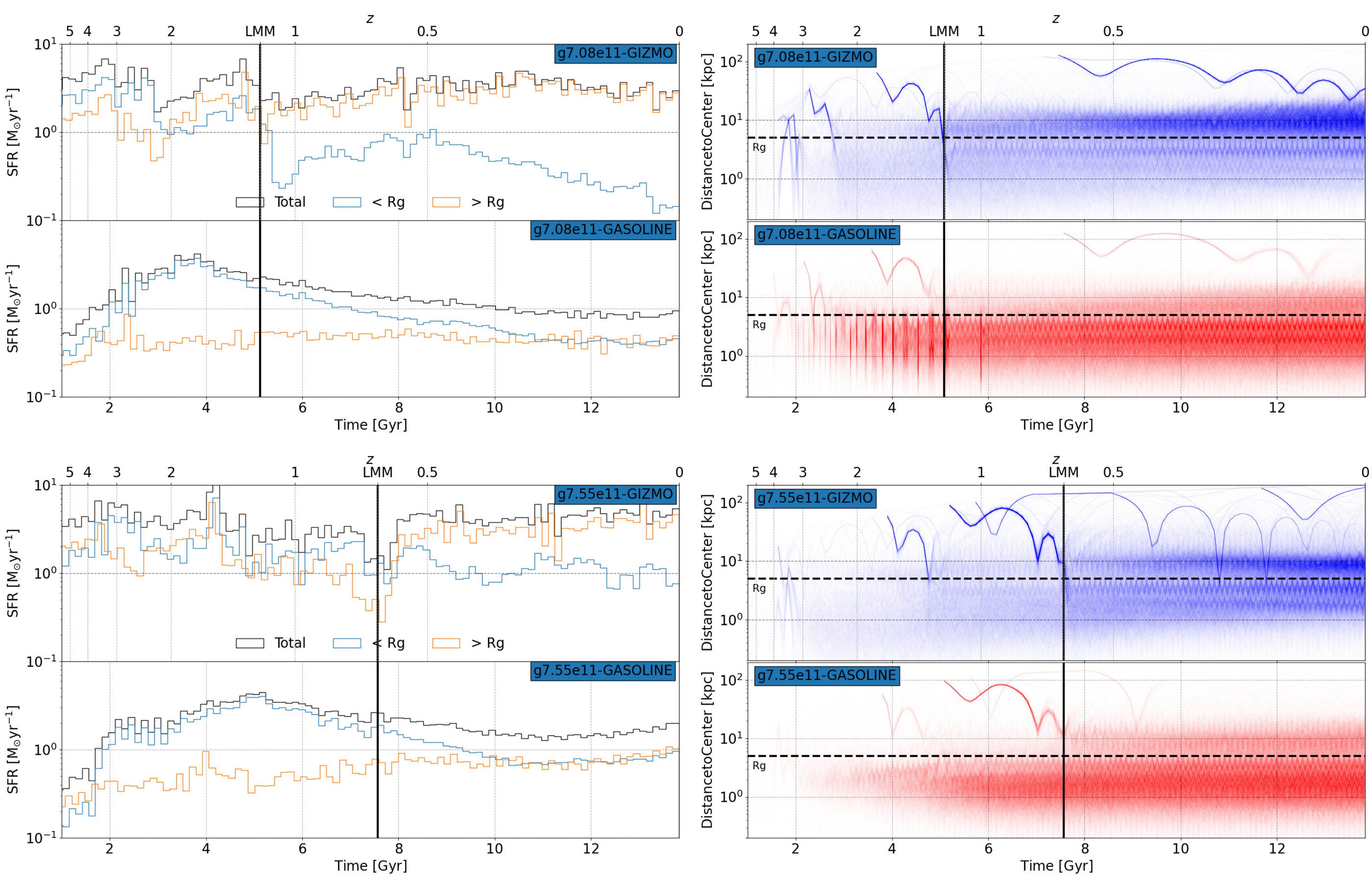}
%\gridline{\fig{sfh_mig.pdf}{0.8\textwidth}}
%\gridline{\fig{sfhr_708.pdf}{0.49\textwidth}{(a)}\fig{sfhr_826.pdf}{0.49\textwidth}{(b)}}
\caption{The star formation histories and the evolution of the radial distance of randomly selected stars in two galaxies (g7.08e11 and g7.55e11). The right panel show the evolution of radial distance to galactic center of 10,000 randomly selected star particles in each galaxy, which clearly show a bimodal distribution with a characteristic radius, $R_g\sim5\rm kpc$, which roughly classify the stars into the inner and outer region of the galaxies. The left panel shows the total SFH, as well as SFHs in the inner and outer region of the galaxies. For a reference, we mark the epoch for the last major merger of each galaxy in each panel.}
\label{fig:sfhr}
\end{figure*}

%is defined within which the volume density $\rho_g$ of the cold ISM is larger than $10^5\,\mathrm M_{\odot}\mathrm{kpc}^{-3}$, as Fig. \ref{fig:hothalo} indicates that both simulations of these systems have a turning point in the cold gas density profile around this value.
In this part we focus on the comparison of star formation histories (SFHs). To distinguish where the star formation happens inside the galaxies, we firstly show the radial distribution of star particles in two galaxies, g7.08e11 and g7.55e11, both with well-developed cold discs at $z=0$, in the right panel of Fig. \ref{fig:sfhr} as examples. Here we randomly selected 10,000 star particles at $z=0$ from each galaxy, and plot their radial distance to galactic center at different redshifts. Note that at high redshifts the number of particles is decreasing as some stellar particles were in gas phase and have not turned into stars, so we do not plot those particles. Firstly, we note that there are a few trajectories in the plot，such as the oscillating blue lines in the upper panel, which are star particles selected from some satellite galaxies, so they orbits are oscillating by passing the peri/apo-center of the host galaxy. It is seen that for most star particles, their distance to the galactic center display a bimodal distribution with a characteristic distance, $R_g\sim5\rm kpc $, which can well separate the star particles into the inner and outer region (see the thick horizontal dashed line in right panels). It is also found that such a bimodal distribution extends to high redshifts, indicating that stars in the galaxies are mainly formed in inner and outer regions respectively, with no clear sign of migration after formation. 

In the left panels of Fig.\ref{fig:sfhr} we show the total SFH of the two galaxies, as well as the SFHs in the inner and outer region divided by the characteristic radius $R_g$. The thick vertical solid line in each panel marks the epoch when the galaxy experiences the last major merger (LMM). The most distinct difference between the total SFH (black line) of galaxy in two simulations is that NIHAO-UHD predicts a decline of SFH with a peak at high redshifts before the LMM, while the SFH derived by \texttt{GIZMO} is more stable across the life time of the galaxy. So the resulting of stellar age derived by \texttt{GIZMO} is younger. In NIHAO-UHD simulation, the early star formation mainly happened in the inner region of the galaxy, such as in the bulge. While in the RiNG simulation, the total SFH before the LMM is more bursty, but with no significant differences in star formation rates within and beyond $R_g$. After the LMM, the star formation appears relatively smooth, driven primarily by star formation in outer/disc regions. In short words, star formation derived by \texttt{GIZMO} is more recently dominated in galactic disc (outer region), but in NIHAO-UHD the recent star formation is equally important in both inner bulge and outer disc regions. Finally, we note that the above difference of SFHs in galaxy inner and outer region will result in different stellar age and metallicity distribution in disc galaxies, which can be further constrained using observational data and we leave it to further study.

%In contrast, SFHs within $R_g$ decline rapidly after LMM, becoming negligible in total SFHs, evidenced by nearly identical orange and black histograms in GIZMO samples at lower redshifts. NIHAO SFHs, however, exhibit subtle indications of the LMM event, with inner star formation dominating total SFHs for most of the time due to the dense core at the center of the NIHAO cold disc (refer to the 4th column of Fig. \ref{fig:fig1}). Additionally, thermal stellar feedback mechanisms in NIHAO lead to an accumulation of hot ISM in the galaxy center (as we have shown in Fig. \ref{fig:phase}). The coexistence of significant amounts of cold and hot ISM is further supported by the NIHAO gas profile illustrated in Fig. \ref{fig:profile}.

\subsection{Origin of the Cold Disc} \label{sec:3.6}

While halo spin can partially account for the properties of discs formed in cosmological simulations \citep{1998MNRAS.295..319M}, our understanding of cold discs, particularly the intricate formation processes of thin disc systems, remains incomplete \citep{2022MNRAS.514.5056H}. In this section we trace the well-developed cold discs formed at $z=0$ in both simulations to higher redshifts, and analyse the mass ratio and angular momentum ratio of their progenitors in each phase. At a specific redshift $z$, we define the angular momentum ratio of phase $i$ as,
\begin{equation}
    a_i(z)=\frac{\boldsymbol{J}_i(z)\boldsymbol{\cdot P}(z)}{\left|\boldsymbol{P}(z)\right|^2},
    \label{equ3}
\end{equation}
where $\boldsymbol{P}$ and $\boldsymbol{J}_i$ represent the total angular momentum of progenitors and their subsets in different spatial and temperature phases, i.e. ISM, CGM, IGM, as well as their combinations with cold, warm and hot. Such a definition naturally guarantees the total is unity,
\begin{equation}
    \sum_ia_i(z)=1.
\end{equation}

\begin{figure*}[htbp]
\gridline{\fig{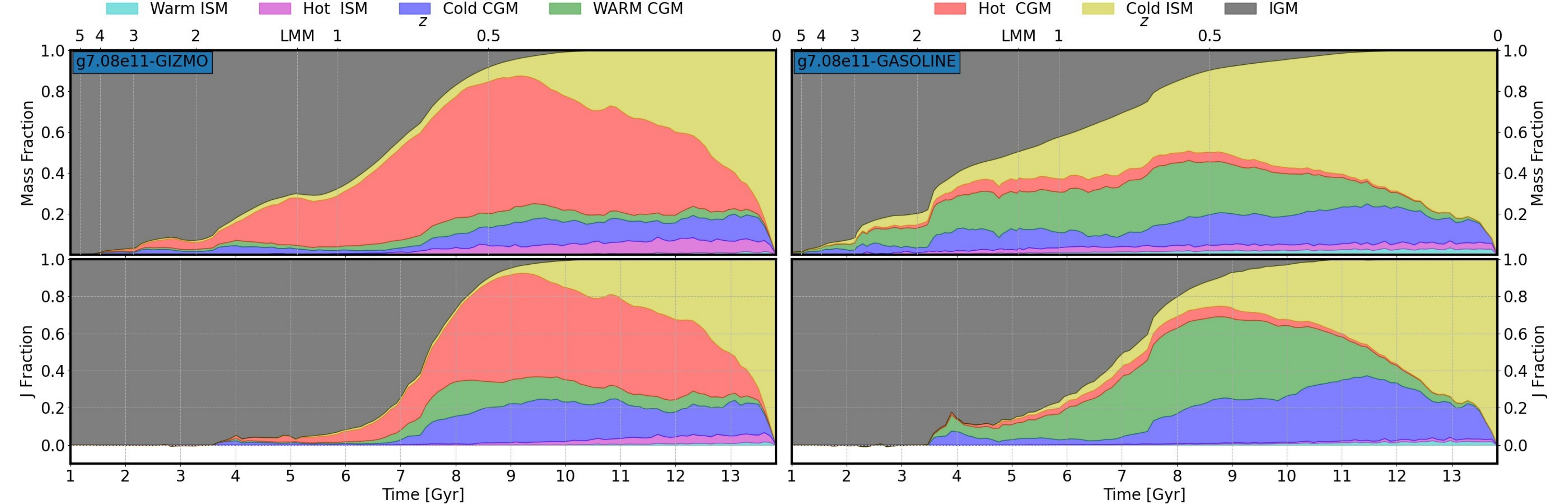}{1.0\textwidth}{(g7.08e11)}}
\gridline{\fig{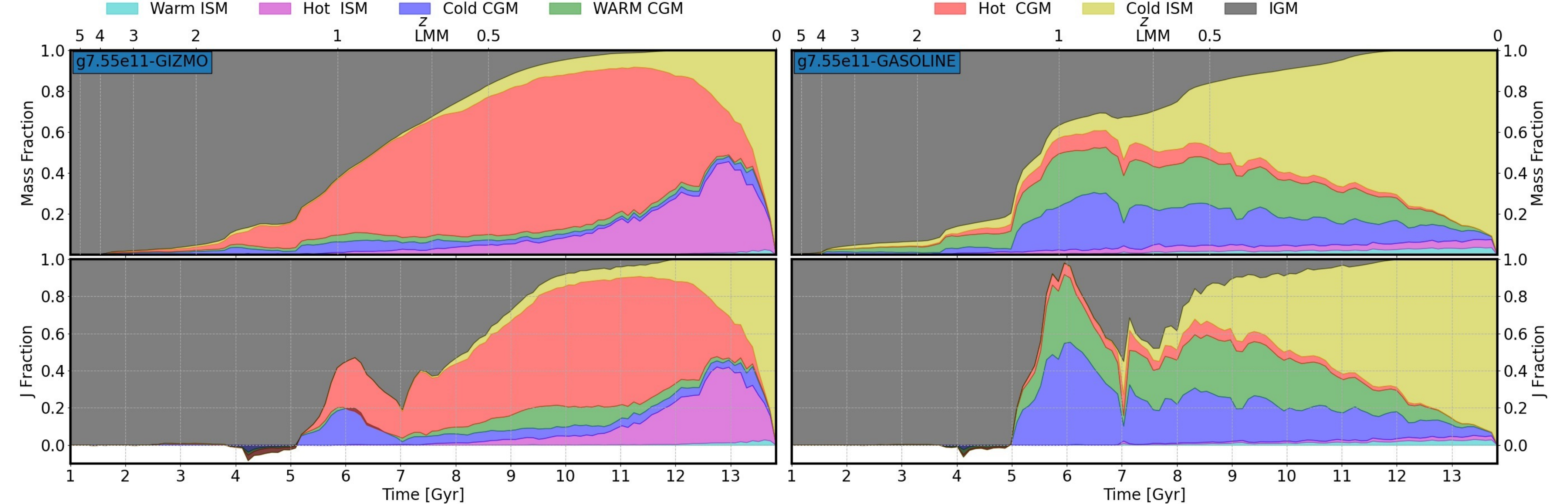}{1.0\textwidth}{(g7.55e11)}}
\caption{Whereabouts of the final cold disc of the g7.08e11 and g7.55e11 galaxy in two simulations across the cosmic time. The upper and lower panels of each subplot display comparisons of the mass and angular momentum fraction in different phases, respectively.}
\label{fig:percent}
\end{figure*}

Fig. \ref{fig:percent} illustrates the g7.08e11 and g7.55e11 galaxy as two examples to highlight the main differences in cold disc formation between the two simulations. It is evident that all gas in the cold disc at $z=0$ originates from the accretion of the IGM at high redshifts. Consequently, the gray and yellow blocks converge towards unity from the left to the right sides, respectively. Throughout the conversion of gas, it passes through different phases, as indicated by the fluctuating colored blocks in the upper four panels. Notably, the most prominent feature in the RiNG simulation is the ``swelling" of the red blocks at a certain redshift, which indicates that after the LMM, gas in the IGM is accreted and began to accumulates as hot CGM for several billion years before transit to the cold (CGM/ISM) phase. Additionally, the accumulated gas in the hot CGM also carries a significant proportion of angular momentum and it gradually transfers from the hot CGM to the cold CGM, and subsequently to the cold disc as the gas cools down. In contrast, the conversion of gas from the IGM to the cold disc in NIHAO-UHD began much earlier. Most of the IGM with a temperature of $10^{4\sim 5}\,\mathrm{K}$ steadily accumulates as warm and cold inflows (refer to the NIHAO-UHD inflow panels in Fig. \ref{fig:inflow}). This warm/cold gas temporarily accumulates in the CGM, experiences rapid loss of pressure, and subsequently falls down to the cold disc. On the other hand, it is found that the mass fraction of un-recycled gas (i.e. which is accreted from the IGM and transferred to the cold disc without undergoing any feedback process) is 30\% in RiNG and 8\% in NIHAO-UHD for this galaxy. Although the un-recycled gas fraction may vary slightly among simulations and galaxy samples, the fact that most of the gas particle in a MW-mass cold disc will experience at least one feedback process before turning into stars has been confirmed by many previous work (for other NIHAO samples please refer to \citet{2019MNRAS.485.2511T}, and also refer to  \citet{2023MNRAS.519.1899C} for recycle fraction comparison between NIHAO and FIRE). Therefore, the transition of most accreted gas from the IGM to the cold disc is not a one-way process, the hot CGM in RiNG simulation and warm/cold CGM in NIHAO-UHD simulation should contain a certain proportion of feedback gas from the cold disc, respectively. This conclusion further enhances our understanding of baryon recycling in the two simulations (see Section \ref{sec:3.4}).

Lastly, we remind readers to take note of the significant fact that gas accreted at high redshifts ($z>2$ for this two systems) exhibits negligible angular momentum, and even reverse in certain instances. This phenomenon is consistent in both two simulations. It indicates that, for MW-mass galaxies, the difference between the final angular momentum of the cold disc and that of the progenitors at $z>2$ may be quite significant.

%In contrast, the gas conversion from IGM to the cold disc in NIHAO began long before GIZMO. A significant amount of IGM with a temperature of $10^{4\sim5}\,\mathrm K$ continuously and steadily accumulates as warm and cold flows (also refer to NIHAO inflow panels in Fig. \ref{fig:inflow}) until late times. The role of hot gas can be neglected. On the other hand, the fraction of fresh accumulated gas that accumulates from IGM to cold disc without experiencing feedback process in cold disc is $30\%$ and $8\%$ in GIZMO and NIHAO, respectively.

%{\bf here you should simple explain how you get the specific J fraction, not in the figure caption}

%This picture is also confirmed by the significant dominance of cold/warm CGM in terms of specific angular momentum after the LMM (lower right panel).  

%It is noteworthy to observe the lower two panels, wherein the substitution of angular momentum with specific one is plotted. It becomes apparent that the specific angular momentum between the hot CGM and ISM exhibits noteworthy similarity in GIZMO.

\section{Discussion and Conclusions} \label{sec:4}

Both NIHAO and FIRE projects are the state-of-art zoom-in hydro-simulation suites and they both have successfully reproduced some observational results for local disc galaxies. However, the different feedback mechanisms implemented in the \texttt{GASOLINE} and \texttt{GIZMO} code could result in significantly different outcomes about the gas distributions. In this paper, we select 12 galaxies from the NIHAO-I and NIHAO-UHD suite, with the halo mass ranges from dwarf ($10^{10}\mathrm{M}_{\odot}$) to MW-mass ($10^{12}\mathrm{M}_{\odot}$), and re-simulate them using the public \texttt{GIZMO} code. The baryonic resolution of these samples is on the order of $\sim10^4\,\mathrm{M}_{\odot}$. In this paper we particularly focus on MW-mass samples which have cold discs formed in both simulations, namely g6.96e11, g7.08e11, g7.55e11, g8.26e11, and g2.79e12, and compare their properties in multiple aspects, such as phase diagram of gas, profiles, SFHs etc. Our main results are summarized as follows.

\begin{itemize}
\item Both \texttt{GASOLINE} and \texttt{GIZMO} codes with the same initial conditions and large-scale environments can successfully produce similar galaxy discs, yet significant differences are still seen in many properties of the galaxies, particularly the CGM environment around them.
\item The early stellar feedback and delayed cooling method employed in \texttt{GASOLINE} effectively suppresses star formation processes, resulting in ubiquitous long-term large-scale outflows primarily driven by the high-density hot ISM from galactic center. In contrast, strong outflows are hardly observed in galaxies simulated by \texttt{GIZMO} code at late cosmic times, but instead display quasi-virialized hot gas halos that arise from the interaction between inflows and feedback-driven outflows. 
\item In the NIHAO-UHD simulations, the cold/warm flows in CGM environment are continuously seen across cosmic time. The mass and angular momentum they carry are crucial for the formation of cold discs. While in the RiNG simulation the formation of cold discs is closely related to the presence of the hot gas halo. The different mechanisms for the formation of cold discs in the two simulations account for the different disc sizes for the same galaxies shown in Fig. \ref{fig:mocks}.
\item Mergers, especially LMM events, are crucial for cold disc formation in galaxies derived by \texttt{GIZMO}. They lead to a sudden deepening of the gravitational well, confining nearly all feedback gas within the virial radius, and transforming bursty SFHs into continuous and smooth ones. In contrast, the impact of merger events on the formation of NIHAO-UHD galaxies is imperceptible.
\end{itemize}

Our study in this paper has revealed that different feedback mechanisms can significantly influence various properties of disc galaxies, particularly the CGM environment in which they are located. Future observational data on the properties of the multi-phase gas components, such as cold/warm CGM inflows, continuous outflows, or quasi-virialized hot gas halos, can be employed to differentiate between the predictions of the \texttt{GASOLINE} and the \texttt{GIZMO}. This can help constrain the different stellar feedback mechanisms in two codes. Additionally, the mock optical images of disc galaxies simulated by \texttt{GIZMO} (refer to Fig. \ref{fig:mocks}) often present more dynamic features, such as bars, bulges, flocculent or grand design arms, merger signatures, etc., compared to the images from the NIHAO suites, where the discs exhibit similarity to each other and lack such features. This contrast suggests that the dynamical properties of the galaxy discs in two simulations are also notably distinct. We will study these dynamical features in future work.

\acknowledgments

We thank the anonymous referee for constructive comments. We acknowledge the support from the National Key Research and Development Program of China (No.2022YFA1602903), the NSFC \ (No. 11825303, 12347103, 11861131006), the science research grants from the China Manned Space project with No. CMS-CSST-2021-A03, CMS-CSST-2021-B01, the Fundamental Research Funds for the Central Universities of China \ (226-2022-00216), and the funds of cosmology simulation database (CSD) in the National Basic Science Data Center (NBSDC). Tobias Buck's contribution to this project was made possible by funding from the Carl Zeiss Stiftung. We also acknowledge the support of the High Performance Computing resources at New York University Abu Dhabi. We gratefully acknowledge the Gauss Centre for Supercomputing e.V. (www.gauss-centre.eu) for funding this project by providing computing time on the GCS Supercomputer SuperMUC at Leibniz Supercomputing Centre (www.lrz.de). Renyue Cen thanks the start-up funding from Zhejiang University.

%\bibliography{sample63}

\begin{thebibliography}{99}

\bibitem[Agertz et al.(2013)]{2013ApJ...770...25A} Agertz, O., Kravtsov, A.~V., Leitner, S.~N., et al.\ 2013, \apj, 770, 25. doi:10.1088/0004-637X/770/1/25
\bibitem[Binney(1977)]{1977ApJ...215..483B} Binney, J.\ 1977, \apj, 215, 483. doi:10.1086/155378
\bibitem[Bird et al.(2022)]{2022MNRAS.516..731B} Bird, S.~A., Xue, X.-X., Liu, C., et al.\ 2022, \mnras, 516, 731. doi:10.1093/mnras/stac2036
\bibitem[Brooks \& Zolotov (2014)]{2014ApJ...786...87B}Brooks, A. \& Zolotov, A.\ 2014, \apj, 786, 87. doi: 10.1088/0004-637X/786/2/87
\bibitem[Buck et al.(2020)]{2020MNRAS.491.3461B} Buck, T., Obreja, A., Macci{\`o}, A.~V., et al.\ 2020, \mnras, 491, 3461. doi:10.1093/mnras/stz3241
\bibitem[Chadayammuri et al.(2022)]{2022ApJ...936L..15C} Chadayammuri, U., Bogd{\'a}n, {\'A}., Oppenheimer, B.~D., et al.\ 2022, \apjl, 936, L15. doi:10.3847/2041-8213/ac8936
\bibitem[Chen et al.(2023)]{2023MNRAS.519.1899C} Chen, Y., Xu, Y., \& Kang, X.\ 2023, \mnras, 519, 1899. doi:10.1093/mnras/stac3628
\bibitem[Dalla Vecchia \& Schaye(2012)]{2012MNRAS.426..140D} Dalla Vecchia, C. \& Schaye, J.\ 2012, \mnras, 426, 140. doi:10.1111/j.1365-2966.2012.21704.x
\bibitem[Di Cintio et al. (2014)]{2014MNRAS.437..415D}Di Cintio, A., et al.\ 2015, \mnras, 437, 415. doi:10.1093/mnras/stt1891
\bibitem[Faucher-Gigu{\`e}re et al.(2009)]{2009ApJ...703.1416F} Faucher-Gigu{\`e}re, C.-A., Lidz, A., Zaldarriaga, M., et al.\ 2009, \apj, 703, 1416. doi:10.1088/0004-637X/703/2/1416
\bibitem[Fitts et al. (2017)]{2017MNRAS.471.3547F}Fitts, A., et al.\ 2017, \mnras, 471, 3547. doi:10.1093/mnras/stx1757
\bibitem[Gaburov \& Nitadori(2011)]{2011MNRAS.414..129G} Gaburov, E. \& Nitadori, K.\ 2011, \mnras, 414, 129. doi:10.1111/j.1365-2966.2011.18313.x
\bibitem[Gill et al.(2004)]{2004MNRAS.351..399G} Gill, S.~P.~D., Knebe, A., \& Gibson, B.~K.\ 2004, \mnras, 351, 399. doi:10.1111/j.1365-2966.2004.07786.x
\bibitem[Governato et al. (2012)]{2012MNRAS.422.1231G}Governato, F., et al.\ 2012, \mnras, 422, 1231. doi:10.1111/j.1365-2966.2012.20696.x
\bibitem[Guszejnov et al.(2017)]{2017MNRAS.472.2107G} Guszejnov, D., Hopkins, P.~F., \& Ma, X.\ 2017, \mnras, 472, 2107. doi:10.1093/mnras/stx2067
\bibitem[Hafen et al.(2022)]{2022MNRAS.514.5056H} Hafen, Z., Stern, J., Bullock, J., et al.\ 2022, \mnras, 514, 5056. doi:10.1093/mnras/stac1603
\bibitem[Hopkins(2015)]{2015MNRAS.450...53H} Hopkins, P.~F.\ 2015, \mnras, 450, 53. doi:10.1093/mnras/stv195
\bibitem[Hopkins et al.(2023)]{2023arXiv230108263H} Hopkins, P.~F., Gurvich, A.~B., Shen, X., et al.\ 2023, arXiv:2301.08263. doi:10.48550/arXiv.2301.08263
\bibitem[Hopkins et al.(2014)]{2014MNRAS.445..581H} Hopkins, P.~F., Kere{\v{s}}, D., O{\~n}orbe, J., et al.\ 2014, \mnras, 445, 581. doi:10.1093/mnras/stu1738
\bibitem[Hopkins et al.(2018)]{2018MNRAS.480..800H} Hopkins, P.~F., Wetzel, A., Kere{\v{s}}, D., et al.\ 2018, \mnras, 480, 800. doi:10.1093/mnras/sty1690
\bibitem[Hopkins et al.(2018)]{2018MNRAS.477.1578H} Hopkins, P.~F., Wetzel, A., Kere{\v{s}}, D., et al.\ 2018, \mnras, 477, 1578. doi:10.1093/mnras/sty674
\bibitem[Hu et al.(2023)]{2023ApJ...950..132H} Hu, C.-Y., Smith, M.~C., Teyssier, R., et al.\ 2023, \apj, 950, 132. doi:10.3847/1538-4357/accf9e
\bibitem[Hudson et al.(2014)]{2014ApJ...787L...5H} Hudson, M.~J., Harris, G.~L., \& Harris, W.~E.\ 2014, \apjl, 787, L5. doi:10.1088/2041-8205/787/1/L5
\bibitem[Kelly et al.(2022)]{2022MNRAS.514.3113K} Kelly, A.~J., Jenkins, A., Deason, A., et al.\ 2022, \mnras, 514, 3113. doi:10.1093/mnras/stac1019
\bibitem[Kere{\v{s}} et al.(2005)]{2005MNRAS.363....2K} Kere{\v{s}}, D., Katz, N., Weinberg, D.~H., et al.\ 2005, \mnras, 363, 2. doi:10.1111/j.1365-2966.2005.09451.x
\bibitem[Kim et al.(2014)]{2014ApJS..210...14K} Kim, J.-. hoon ., Abel, T., Agertz, O., et al.\ 2014, \apjs, 210, 14. doi:10.1088/0067-0049/210/1/14
\bibitem[Kimm \& Cen(2014)]{2014ApJ...788..121K} Kimm, T. \& Cen, R.\ 2014, \apj, 788, 121. doi:10.1088/0004-637X/788/2/121
\bibitem[Kimm et al.(2015)]{2015MNRAS.451.2900K} Kimm, T., Cen, R., Devriendt, J., et al.\ 2015, \mnras, 451, 2900. doi:10.1093/mnras/stv1211
\bibitem[Knollmann \& Knebe(2009)]{2009ApJS..182..608K} Knollmann, S.~R. \& Knebe, A.\ 2009, \apjs, 182, 608. doi:10.1088/0067-0049/182/2/608
\bibitem[Lazar et al. (2020)]{2020MNRAS.497.2393L}Lazar, A., et al.\ 2020, \mnras, 497, 2392. doi:10.1093/mnras/staa2101
\bibitem[Leitherer et al.(1999)]{1999ApJS..123....3L} Leitherer, C., Schaerer, D., Goldader, J.~D., et al.\ 1999, \apjs, 123, 3. doi:10.1086/313233
\bibitem[Lelli et al.(2016)]{2016AJ....152..157L} Lelli, F., McGaugh, S.~S., \& Schombert, J.~M.\ 2016, \aj, 152, 157. doi:10.3847/0004-6256/152/6/157
\bibitem[Lim et al.(2021)]{2021MNRAS.504.5131L} Lim, S.~H., Barnes, D., Vogelsberger, M., et al.\ 2021, \mnras, 504, 5131. doi:10.1093/mnras/stab1172
\bibitem[Macci{\`o} et al.(2012)]{2012ApJ...744L...9M} Macci{\`o}, A.~V., Stinson, G., Brook, C.~B., et al.\ 2012, \apjl, 744, L9. doi:10.1088/2041-8205/744/1/L9
\bibitem[Macci{\`o} et al.(2016)]{2016MNRAS.463L..69M} Macci{\`o}, A.~V., Udrescu, S.~M., Dutton, A.~A., et al.\ 2016, \mnras, 463, L69. doi:10.1093/mnrasl/slw147
\bibitem[Macci{\`o} et al.(2020)]{2020MNRAS.495L..46M} Macci{\`o}, A.~V., Crespi, S., Blank, M., et al.\ 2020, \mnras, 495, L46. doi:10.1093/mnrasl/slaa058
\bibitem[McMillan(2011)]{2011MNRAS.414.2446M} McMillan, P.~J.\ 2011, \mnras, 414, 2446. doi:10.1111/j.1365-2966.2011.18564.x
\bibitem[Mo et al.(1998)]{1998MNRAS.295..319M} Mo, H.~J., Mao, S., \& White, S.~D.~M.\ 1998, \mnras, 295, 319. doi:10.1046/j.1365-8711.1998.01227.x
\bibitem[Muratov et al.(2015)]{2015MNRAS.454.2691M} Muratov, A.~L., Kere{\v{s}}, D., Faucher-Gigu{\`e}re, C.-A., et al.\ 2015, \mnras, 454, 2691. doi:10.1093/mnras/stv2126
\bibitem[Naab \& Ostriker(2017)]{2017ARA&A..55...59N} Naab, T. \& Ostriker, J.~P.\ 2017, \araa, 55, 59. doi:10.1146/annurev-astro-081913-040019
\bibitem[Navarro et al.(1995)]{1995MNRAS.275...56N} Navarro, J.~F., Frenk, C.~S., \& White, S.~D.~M.\ 1995, \mnras, 275, 56. doi:10.1093/mnras/275.1.56
\bibitem[Navarro et al.(1996)]{1996MNRAS.283L..72N} Navarro, J.~F., Eke, V.~R., \& Frenk, C.~S.\ 1996, \mnras, 283, L72. doi:10.1093/mnras/283.3.L72
\bibitem[Navarro \& Steinmetz(1997)]{1997ApJ...478...13N} Navarro, J.~F. \& Steinmetz, M.\ 1997, \apj, 478, 13. doi:10.1086/303763
\bibitem[Pakmor et al.(2016)]{2016MNRAS.455.1134P} Pakmor, R., Springel, V., Bauer, A., et al.\ 2016, \mnras, 455, 1134. doi:10.1093/mnras/stv2380
\bibitem[Peebles(1969)]{1969ApJ...155..393P} Peebles, P.~J.~E.\ 1969, \apj, 155, 393. doi:10.1086/149876
\bibitem[Pillepich et al.(2018)]{2018MNRAS.473.4077P} Pillepich, A., Springel, V., Nelson, D., et al.\ 2018, \mnras, 473, 4077. doi:10.1093/mnras/stx2656
\bibitem[Planck Collaboration et al.(2014)]{2014A&A...571A..16P} Planck Collaboration, Ade, P.~A.~R., Aghanim, N., et al.\ 2014, \aap, 571, A16. doi:10.1051/0004-6361/201321591
\bibitem[Pontzen et al.(2013)]{2013ascl.soft05002P} Pontzen, A., Ro{\v{s}}kar, R., Stinson, G., et al.\ 2013, Astrophysics Source Code Library. ascl:1305.002
\bibitem[Rees \& Ostriker(1977)]{1977MNRAS.179..541R} Rees, M.~J. \& Ostriker, J.~P.\ 1977, \mnras, 179, 541. doi:10.1093/mnras/179.4.541
\bibitem[Richings et al.(2014b)]{2014MNRAS.442.2780R} Richings, A.~J., Schaye, J., \& Oppenheimer, B.~D.\ 2014, \mnras, 442, 2780. doi:10.1093/mnras/stu1046
\bibitem[Richings et al.(2014a)]{2014MNRAS.440.3349R} Richings, A.~J., Schaye, J., \& Oppenheimer, B.~D.\ 2014, \mnras, 440, 3349. doi:10.1093/mnras/stu525
\bibitem[Santos-Santos et al.(2018)]{2018MNRAS.473.4392S} Santos-Santos, I.~M., Di Cintio, A., Brook, C.~B., et al.\ 2018, \mnras, 473, 4392. doi:10.1093/mnras/stx2660
\bibitem[Scannapieco et al.(2006)]{2006MNRAS.371.1125S} Scannapieco, C., Tissera, P.~B., White, S.~D.~M., et al.\ 2006, \mnras, 371, 1125. doi:10.1111/j.1365-2966.2006.10785.x
\bibitem[Scannapieco et al.(2012)]{2012MNRAS.423.1726S} Scannapieco, C., Wadepuhl, M., Parry, O.~H., et al.\ 2012, \mnras, 423, 1726. doi:10.1111/j.1365-2966.2012.20993.x
\bibitem[Schaye et al.(2015)]{2015MNRAS.446..521S} Schaye, J., Crain, R.~A., Bower, R.~G., et al.\ 2015, \mnras, 446, 521. doi:10.1093/mnras/stu2058
\bibitem[Schaye \& Dalla Vecchia(2008)]{2008MNRAS.383.1210S} Schaye, J. \& Dalla Vecchia, C.\ 2008, \mnras, 383, 1210. doi:10.1111/j.1365-2966.2007.12639.x
\bibitem[Springel(2005)]{2005MNRAS.364.1105S} Springel, V.\ 2005, \mnras, 364, 1105. doi:10.1111/j.1365-2966.2005.09655.x
\bibitem[Springel \& Hernquist(2003)]{2003MNRAS.339..289S} Springel, V. \& Hernquist, L.\ 2003, \mnras, 339, 289. doi:10.1046/j.1365-8711.2003.06206.x
\bibitem[Stinson et al.(2013)]{2013MNRAS.428..129S} Stinson, G.~S., Brook, C., Macci{\`o}, A.~V., et al.\ 2013, \mnras, 428, 129. doi:10.1093/mnras/sts028
\bibitem[Stinson et al.(2006)]{2006MNRAS.373.1074S} Stinson, G., Seth, A., Katz, N., et al.\ 2006, \mnras, 373, 1074. doi:10.1111/j.1365-2966.2006.11097.x
\bibitem[Teyssier(2002)]{2002A&A...385..337T} Teyssier, R.\ 2002, \aap, 385, 337. doi:10.1051/0004-6361:20011817
\bibitem[Tollet et al.(2016)]{2016MNRAS.456.3542T} Tollet, E., Macci{\`o}, A.~V., Dutton, A.~A., et al.\ 2016, \mnras, 456, 3542. doi:10.1093/mnras/stv2856
\bibitem[Tollet et al.(2019)]{2019MNRAS.485.2511T} Tollet, {\'E}., Cattaneo, A., Macci{\`o}, A.~V., et al.\ 2019, \mnras, 485, 2511. doi:10.1093/mnras/stz545
\bibitem[Wadsley et al.(2004)]{2004NewA....9..137W} Wadsley, J.~W., Stadel, J., \& Quinn, T.\ 2004, \na, 9, 137. doi:10.1016/j.newast.2003.08.004
\bibitem[Wang et al.(2015)]{2015MNRAS.454...83W} Wang, L., Dutton, A.~A., Stinson, G.~S., et al.\ 2015, \mnras, 454, 83. doi:10.1093/mnras/stv1937



\end{thebibliography}
%\bibliographystyle{aasjournal}

\end{CJK*}
\end{document}